%% file: se-ext.tex
\documentclass[11pt,a4paper]{article}
\usepackage[british]{babel}
\usepackage{a4wide,aa-style,enumerate}
\usepackage[T1]{fontenc}
\usepackage[dvips]{graphicx}
\usepackage{psfrag,times}
\input{aa-macros}

\newcommand{\gen}{\tilde{f}}

\newcommand{\rvGamma}{\mathrm{Gamma}}
\newcommand{\St}{\mathop{\mathrm{St}}\nolimits}
\newcommand{\PVII}{\mathop{\mathrm{PVII}}\nolimits}
\newcommand{\PII}{\mathop{\mathrm{PII}}\nolimits}
\newcommand{\Ell}{\mathop{\mathrm{Ell}}\nolimits}
\title{
       Distributions generated by perturbation of symmetry \\
       with emphasis on a multivariate skew $t$ distribution}
\author{
     \textsc{Adelchi Azzalini} \\
       {Dipartimento di Scienze Statistiche,  Università di Padova} \\
       \texttt{azzalini@stat.unipd.it}
   \and
        \textsc{Antonella Capitanio} \\
       {Dipartimento di Scienze Statistiche,  Università di Bologna}  \\
       \texttt{capitani@stat.unibo.it}
  }
\date{23rd September 2002\\
(Original version April 2001, last amendement 27th June 2003)\\
~\\
\textsl{This is the full-length paper whose abriged version appears in\\
   \emph{J.\,Roy.\,Statist.\,Soc., series B} vol.\,65 (2003), pp.\,367--389
}}

\begin{document}
\maketitle
\begin{abstract}

  A fairly general procedure is studied to perturbate a multivariate
  density satisfying a weak form of multivariate symmetry, and to
  generate a whole set of non-symmetric densities. The approach is general
  enough to encompass a number of recent proposals in the literature,
  variously related to the skew normal distribution.
  The special case of skew elliptical densities is examined in
  detail, establishing connections with existing similar work.
  The final part of the paper specializes further to a form of
   multivariate skew $t$ density. Likelihood inference for this
  distribution is examined, and it is illustrated with numerical
  examples.

\end{abstract}
\vspace{3ex} \noindent\emph{Key-words:~} asymmetry,  
central symmetry, elliptical distributions, Healy's plot, 
multivariate $t$ distribution,   quadratic forms,
skewness, skew normal  distribution.

\clearpage

\section{Introduction}

\subsection{Motivation and aims}

There is a growing interest in the literature on parametric
families of multivariate distributions which represent a local
departure from the multivariate normal family, in the sense that
they exhibit a bell-shaped behaviour similar to the normal
density, and they can be made arbitrarily close to the normal
density by regulating a suitable parameter. The  phrase `local
departure' must be interpreted appropriately, in the sense that,
while these families can approach normality, they  also can, under
other circumstances, exhibit quite a substantial departure from
normality.

The motivation of these efforts is to introduce more flexible
parametric families capable of adapting as closely as possible to
real data, in particular in the rather frequent case of phenomena
whose empirical outcome behaves in a non-normal fashion but still
retains some broad similarity with the multivariate normal
distribution. Typically this departure from normality occurs in
the form of a roughly bell-shaped density, but with contour levels
not quite elliptically shaped and/or with contour levels not quite
spaced as the normal density prescribes.

Some of this literature is connected with the so-called
multivariate skew normal (SN) distribution, recently studied by
Azzalini \& Dalla Valle (1996) and Azzalini \& Capitanio (1999);
this has been further developed by other authors whose work will
be referenced later in this section. The $d$-dimensional SN
density, in the `standard' form which does not include location
and scale parameters, is
\begin{equation}  \label{f:sn0}
      2 \: \phi_d(y;\bar\Omega)\:\Phi(\alpha\T y), \qquad y\in\Real^d,
\end{equation}
where $\phi_d(y;\bar\Omega)$ is the $N_d(0,\bar\Omega)$ density at
$y$ for some correlation matrix $\bar\Omega$,  $\Phi(\cdot)$ is
the $N(0,1)$ distribution function and $\alpha\in\Real^d$. Here
$\alpha$ plays the role of shape parameter; when $\alpha=0$, we
recover  the regular normal density.

As a further level of generalisation of the normal distribution,
Azzalini \& Capitanio (1999, p.\,599) have presented a lemma which
leads to the construction of a `skew elliptical' density, which is
an elliptical density multiplied by a suitable skewing factor, in
such a way that the product is still a proper density. Branco \&
Dey (2001) have considered another form of skew elliptical
distribution, whose connections with  the one mentioned above will
be discussed extensively in this paper.  Other work on extensions
of elliptical families has been done by Genton \& Loperfido (2002),
where it is shown that distributional properties of certain functions 
of elliptical variates extends to their skewed variants, generalizing
a similar result of Branco \&Dey (2001).

Arnold \& Beaver (2000a) have studied a variant of (\ref{f:sn0})
which replaces  the argument of $\Phi$ by  $\alpha_0+\alpha\T y$,
where $\alpha_0$ is an additional parameter,  with consequent
adjustment of the normalising constant.  The same variant of the
SN distribution has been considered by Capitanio \emph{et al.}
(2003) in the context of graphical models. Sahu, Dey \& Branco
(2001) have studied yet another form of skew elliptical
distribution, where the skewing factor is a $d$-dimensional
distribution function, rather than a scalar one like those of the
previously mentioned cases. In the same spirit as (\ref{f:sn0}),
Arnold \& Beaver (2000b) have studied  a form of multivariate skew
Cauchy distribution. For additional references and a recent review 
on the connected literature, see Arnold \& Beaver (2002).

There is therefore a set of interesting developments in various
directions aimed at extending (\ref{f:sn0}) or adapting the
underlying idea to other distributions. While all this activity is
definitely promising and appealing, it also brings in the question
of the inter-relationships among these contributions, which tend
to appear as scattered in different  directions.

One purpose of the present contribution is to propose a fairly
general extension of (\ref{f:sn0}); in addition, a better
understanding of the connections and similarities among some of
the above-described proposals is attempted. A broad formulation is
presented in Section~2, and is specialised to a skew elliptical
form in Section~3. This approach encompasses several of the
existing proposals and  it appears to provide a potentially
general framework for special cases. We discuss in some detail a
few of these and, from Section~4 onwards, we focus on a form of
multivariate skew $t$ distribution; since this represents a
mathematically quite  manageable distribution, allowing ample
flexibility in skewness  and kurtosis, it appears to be a
promising tool for a wide range of practical problems. Associated
likelihood inference for this skew $t$ distribution and
illustrative examples are presented in Section~5. Some background
information on the SN distribution and the elliptical family is
given in the second part of this introductory section.

\subsection{Some preliminaries}  \label{s:prelim}

\paragraph{The SN distribution}
Given a full-rank $d\times d$ covariance matrix
$\Omega=(\omega_{rs})$, define
\[
     \omega   = \diag(\omega_1,\dots,\omega_d)
              = \diag(\omega_{11},\dots,\omega_{dd})^{1/2}
\]
and let $\bar\Omega=\omega\inv\Omega\omega\inv$ be the associated
correlation matrix; also let  $\xi, \,\alpha\in\Real^d$. A
$d$-dimen\-sional random variable $Z$ is said to have a skew
normal distribution if it is continuous with density function at
$z\in\Real^d$ of type
\begin{equation}  \label{f:sn}
    2 \,\phi_d(z-\xi;\Omega)\,\Phi(\alpha\T\omega\inv(z-\xi)).
\end{equation}
We shall then write $Z\sim\SN_d(\xi,\Omega,\alpha)$, referring to
$\xi, \Omega, \alpha$ as the location, dispersion and shape or
skewness parameters, respectively. Density (\ref{f:sn0})
corresponds to the `standard' distribution
$\SN_d(0,\bar\Omega,\alpha)$.

By varying $\alpha$, one obtains a variety of shapes; Azzalini \&
Dalla Valle (1996) display graphically some instances of them when
$d=2$. Clearly, when $\alpha=0$, we are back to the
$\N_d(\xi,\Omega)$ density. The cumulant generating function is
\[
   K_Z(t) = t\T \xi +\half t\T \Omega t +\zeta_0(\delta\T \omega t)
\]
where
\begin{equation}  \label{f:delta-zeta}
  \delta = \frac{1}{\radice{1+\alpha\T \bar\Omega\alpha}} \bar\Omega\alpha,
  \qquad
  \zeta_0(x)=\log\{2\,\Phi(x)\} .
\end{equation}
From the expression for $\delta$ we have
\begin{equation}  \label{f:alpha}
  \alpha= \frac{1}
       {\radice{1-\delta\T\bar\Omega\inv\delta}}\:\bar\Omega\inv\delta.
\end{equation}

There exists at least two stochastic representations for $Z$.
These are useful for random number generation and for deriving in
a simple way a number of formal properties.
\begin{itemize}
\item
  \emph{Conditioning method}.  Suppose that $U_0$
  is a scalar random variable and $U$ is a $d$-dimensional variable,
  such that
  \begin{equation}
      \pmatrix{U_0\cr U} \sim \N_{d+1}\left(0, \Omega^* \right) \,,
        \qquad
      \Omega^* = \pmatrix{ 1 & \delta\T \cr
                      \delta & \bar\Omega }
  \label{f:sr-condit}
  \end{equation}
where $\Omega^*$ is a full-rank correlation matrix. Then the
distribution of $(U|U_0>0)$ is $\SN_d(0,\bar\Omega,\alpha)$ where
$\alpha$ is a function of $\delta$  and $\bar\Omega$; in fact, we
can also set
\[
    Z =\cases{U & if  $U_0>0$, \cr
             -U & if  $U_0<0$. }
\]
By an affine transformation of the resulting variable one obtains
a distribution of type (\ref{f:sn}). \item
  \emph{Transformation method}. Suppose now that
\begin{equation}  \label{f:Omega^*}
  \pmatrix{U'_0\cr U'} \sim
       \N_{d+1}\left(0, \pmatrix{ 1 & 0 \cr
                     0 & \Psi } \right)
\end{equation}
where $\Psi$ is a full-rank correlation matrix, and define
\begin{equation} \label{f:sr-transf}
    Z_j = \delta_j\: |U'_0| + \radice{1-\delta_j^2}\: U'_j,
\end{equation}
where $-1<\delta_j<1$ for $j=1,\dots,d$. Then $(Z_1,\dots,Z_d)$ has
the $d$-dimensional skew normal distribution, with parameters
which are suitable functions of the $\delta$'s and $\Psi$.
\end{itemize}

A third type of representation is known to exist in the scalar
case. If $(U_0,U_1)$ is a bivariate normal variate with
standardized marginals and correlation $\rho$, then
\begin{equation} \label{f:max}
         \max(U_0,U_1) \sim~\SN(0,1,\alpha)
\end{equation}
where $\alpha=\radice{(1-\rho)/(1+\rho)}$.  This result has been given
by Roberts (1966), in an early explicit occurrence of the scalar SN
distribution, and later rediscovered by Loperfido (2002); the same
conclusion can also be obtained as special case of a result of H.~N.\
Nagaraja, quoted by David (1981, Exercise 5.6.4).  The generalization
of this type of representation to the multivariate setting to obtain 
(\ref{f:sn0}) via a set of $\max(\cdot)$ operation on normal variates 
is an open question.

Among the many formal properties shared with the normal class, a
noteworthy fact is that
\begin{equation}  \label{f:chi2}
     (Z-\xi)\T \Omega\inv (Z-\xi) \sim \chi^2_d.
\end{equation}
Other properties of quadratic forms of SN variables are given by
Azzalini \& Capitanio (1999),  Genton \emph{et al.} (2001) and
Loperfido (2001). Another important property of this class is
closure under  affine transformations of the variable Z; in
particular, this implies closure under marginalization, i.e.\ the
distribution of all sub-vectors of $Z$ is still of type
(\ref{f:sn}).

What is lacking is closure under conditioning, i.e. the
conditional distribution of a set of components of $Z$ given
another set of components is not of type (\ref{f:sn}). This
property is achieved by a simple extension  of (\ref{f:sn}) which
has been examined by Arnold \& Beaver (2000a) and by Capitanio
\emph{et al.} (2003). This variant of the density takes the form
\begin{equation}  \label{f:esn}
    \Phi(\tau)\inv \,\phi_d(z-\xi;\Omega)\,
                   \Phi(\alpha_0+\alpha\T\omega\inv(z-\xi))
\end{equation}
where $\tau \: (\tau\in\Real)$ is an additional parameter and
\[
 \alpha_0 = \recradice{1-\delta\T \bar\Omega\inv\delta} \tau\,.
\]
When $\tau=0$, $\alpha_0=0$ and (\ref{f:esn}) reduces to
(\ref{f:sn}).  Unfortunately, the $\chi^2$ property (\ref{f:chi2})
does not hold for (\ref{f:esn}), if $\tau\not=0$. A form of
genesis of  (\ref{f:esn}) via conditioning using (\ref{f:Omega^*})
is by consideration of $(U|U_0+\tau>0)$ .


\paragraph{Elliptical distributions}
We summarize briefly a few concepts about and establish notation
for elliptical distributions, confining ourselves to random
variables without discrete components. For a full treatment of
this topic, we refer the reader to Fang, Kotz and Ng (1990).

A $d$-dimensional continuous random variable $Y$ is said to have
an elliptical density if this is of the form
\[
  f(y; \xi, \Omega) = \frac{c_d}{|\Omega|^{1/2}}\:
           \gen\{(y-\xi)\T\Omega\inv (y-\xi)\}, \qquad y\in\Real^d,
\]
where $\xi\in\Real^d,$ $\Omega$ is a covariance matrix, $\gen$ is
a suitable function from $\Real^+$ to $\Real^+$, called the
`density generator', and $c_d$ is a normalising constant. We shall
then write $Y \sim \Ell_d(\xi,\Omega, \gen)$.

The basic case is obtained by setting $\gen(x)=\exp(- x/2)$ and
$c_d=(2\pi)^{-d/2}$, leading to
 the multivariate normal density.
Two other important special cases, which will be used extensively
in the sequel, are provided by the multivariate Pearson type VII
distributions, whose generator and normalising constant are
\[
  \gen(x)= (1+x/\nu)^{-M},  \qquad
   c_d= \frac{\Gamma(M)}{(\pi\nu)^{d/2}\,\Gamma(M-d/2)},
\]
where $\nu>0,M>d/2$, and by  the multivariate Pearson type II
distributions for which
\[
   \gen(x)= (1-x)^\nu,  \qquad
   c_d= \frac{\Gamma(d/2+\nu+1)}{\pi^{d/2}\,\Gamma(\nu+1)}
\]
where $0\leq x \leq 1, \nu>-1$. The special importance  of type
VII lies in the fact that it includes the multivariate $t$ density
when $M=(d+\nu)/2$, hence also the Cauchy distribution. For these
distributions,  we shall use the notation
$\PVII_d(\xi,\Omega,M,\nu)$ and $\PII_d(\xi,\Omega,\nu)$,
respectively.

A convenient stochastic representation for $Y$ is
\begin{equation}    \label{f:ru}
  Y = \xi + R L\T S
\end{equation}
where $L\T L=\Omega$, the random vector $S$ is uniformly
distributed on the unit sphere in $\Real^d$ and $R$ is a positive
scalar random variable independent of $S$, called the generating
variate. An immediate consequence of this representation is that
$(Y-\xi)\T \Omega\inv (Y-\xi) \equald R^2$, where $\equald$ means
equality in distribution.

Elliptical distributions are closed under affine transformations
and conditioning. In particular they are closed under
marginalization, in the following sense:  consider the  block
partition $Y\T = (Y_1\T,Y_2\T)$ where $Y_1 \in \Real^h$ and a
corresponding partition for $\xi$ and $\Omega$; then
\[
    Y_1 \sim \Ell_h(\xi_1, \Omega_{11}, \gen_1)
\]
Similarly, for the conditional density we have
\[ (Y_1|Y_2=y_2) \sim
  \Ell_h(\xi_1 +  \Omega_{12}\Omega_{22}\inv (y_2 - \xi_2),
      \Omega_{11}-\Omega_{12}\Omega_{22}\inv\Omega_{21},\gen^{Q_y}).
\]
where $Q_y= y_2\T\Omega_{22}\inv y_2$. The density generators
$\gen_1$ and $\gen^{Q_y}$ are not necessarily of the same form as
$\gen$. Kano (1994) has shown that the form of the density
generator is preserved under marginalization only in the case of
elliptical distributions which can be obtained from a scale
mixture of normal variates. This property is true, for instance,
for multivariate Pearson type VII and II distributions. The
generator $\gen^{Q_y}$ of the conditional distribution depends in
general on the quantity $Q_y$, with the exception of the normal
distribution.


\section{Central symmetry and distributions obtained by its perturbation}
\label{s:c-symmetry}

Our starting point is the following proposition which is closely
connected to Lemma~1 of Azzalini \& Capitanio (1999). Strictly
speaking, the present statement is a bit more restricted than the
earlier result,  but it has the major advantage of requiring a set
of conditions whose fulfillment is far simpler to check, and still
it represents a very general formulation.

The result refers to central symmetry, a simple and wide concept of
symmetry, which is commonly in use in nonparametric statistics; see
Zuo \& Serfling (2000). Other authors refer to the same property with
alternative terms. A $d$-dimensional random variable $Y$ is said to be
centrally symmetric around a point $\xi$ if $Y-\xi\equald\xi-Y$. Since
we shall deal with continuous variables, the above requirement implies
that the corresponding density function $f$ satisfies
$f(y-\xi)=f(\xi-y)$ for all $y\in\Real^d$, up to a negligible set.  
It is immediate to see that the condition of central symmetry is
satisfied by various ample families, notably the elliptical densities,
but also many others; some examples are the symmetric stable laws, the
Watson rotational symmetric densities, the class of distributions
studied studied by Szab{\l}owski (1998), among many others.

\begin{proposition} \label{p:df-se}
Denote by $f(y)$ the density function of a $d$-dimensional
continuous random variable  which is centrally symmetric around 0,
and by $G$ a scalar distribution function such that $G(-x)=1-G(x)$
for all real $x$. If $w(y)$ is a function from $\Real^d$ to
$\Real$ such that $w(-y)=-w(y)$ for all $y\in\Real^d$, then
\begin{equation}  \label{f:pdf}
   2 \: f(y)\:G\{w(y)\}
\end{equation}
is a density function.
\end{proposition}
\emph{Proof.} Denote by $Y$ a random variable with density $f$,
and by $X$ a random variable with distribution function $G$,
independent of $Y$. To show that $W=w(Y)$ has a distribution
symmetric about 0, denote by $A$  a Borel set of the real line and
by $-A$ its mirror set obtained by  reversing the sign of each
element of $A$. Then, taking into account that $Y$ and $-Y$ have
the same distribution,
\[
   \pr{W\in -A} = \pr{-W \in A}
                = \pr{w(-Y)\in A}
                = \pr{w(Y) \in A},
\]
showing that $W$ has the property indicated. Then, on noticing
that $X-W$ has distribution symmetric about 0, write
\[ \half = \pr{X\leq W} = \E[Y]{\pr{X\leq w(Y)|Y=y}}
         = \int_{\Real^d} G\{w(y)\}\,f(y)\d{y}  \]
which completes the proof.

To demonstrate graphically the ample flexibility attained by
(\ref{f:pdf}) for appropriate choices of $f$, $G$, and $w$,
we present the following example in the case $d=2$. Consider the
non-elliptical distribution
\[
  f(y) = \frac{(1-y_1^2)^{a-1}\:(1-y_2^2)^{b-1}}
           {4^{a+b-1}\,B(a,a)\,B(b,b)}, \qquad y=(y_1,y_2)\in(-1,1)^2,
\]
obtained by multiplication of two symmetric Beta densities rescaled 
to the interval $(-1,1)$, with positive parameters $a$ and $b$. We
perturb this density by choosing
\[
    G(x) = \frac{e^x}{1+e^x}, \qquad
    w(y)= \frac{\sin(p_1 y_1+p_2 y_2)}{1+\cos(q_1 y_1+ q_2 y_2)}
\]
where $p_1, p_2, q_1$ and $q_2$ are additional parameters. 

\begin{figure}
\centerline{
  \includegraphics[width=0.50\hsize,height=68.5mm]{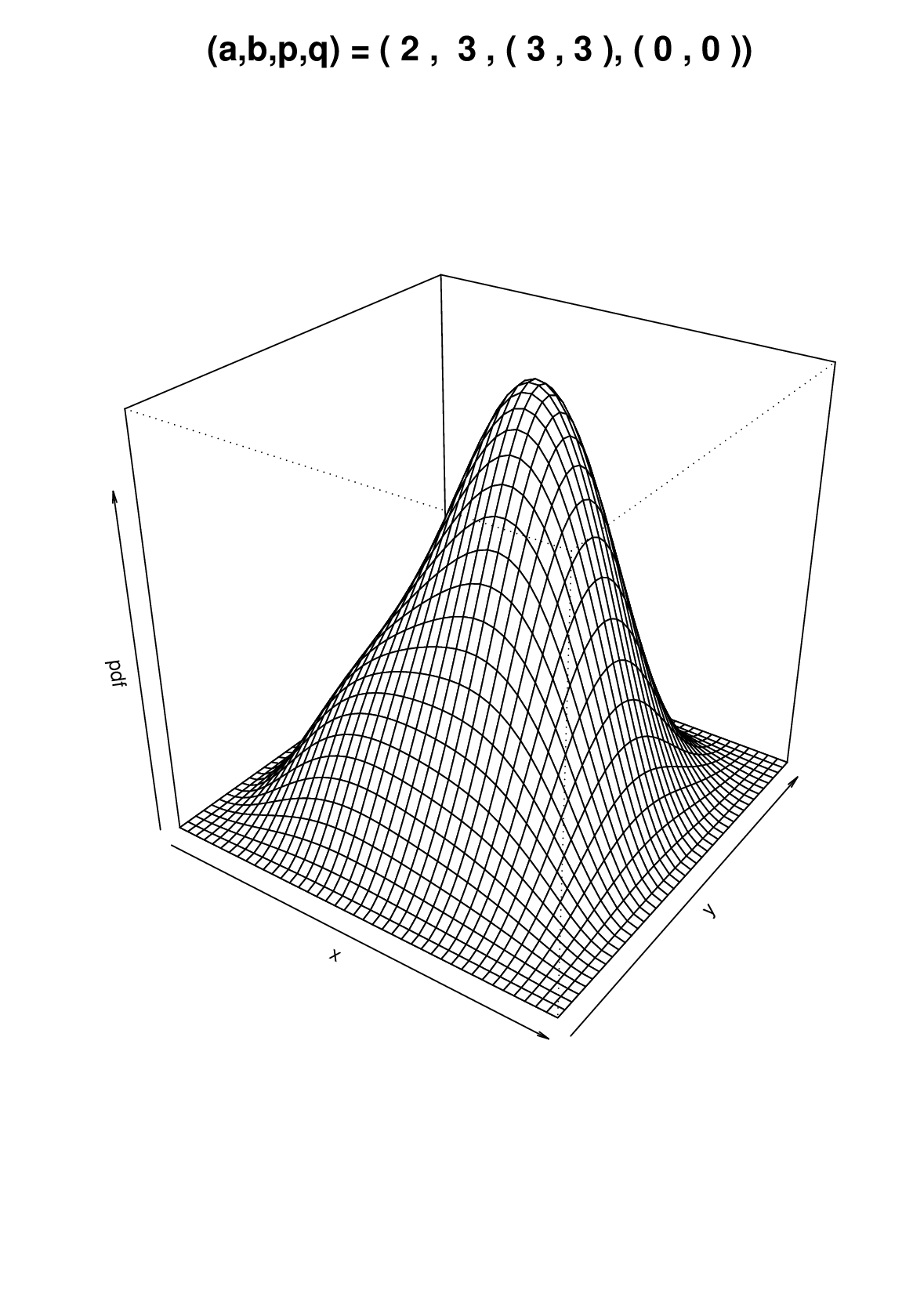}
  \,
  \includegraphics[width=0.50\hsize,height=68.5mm]{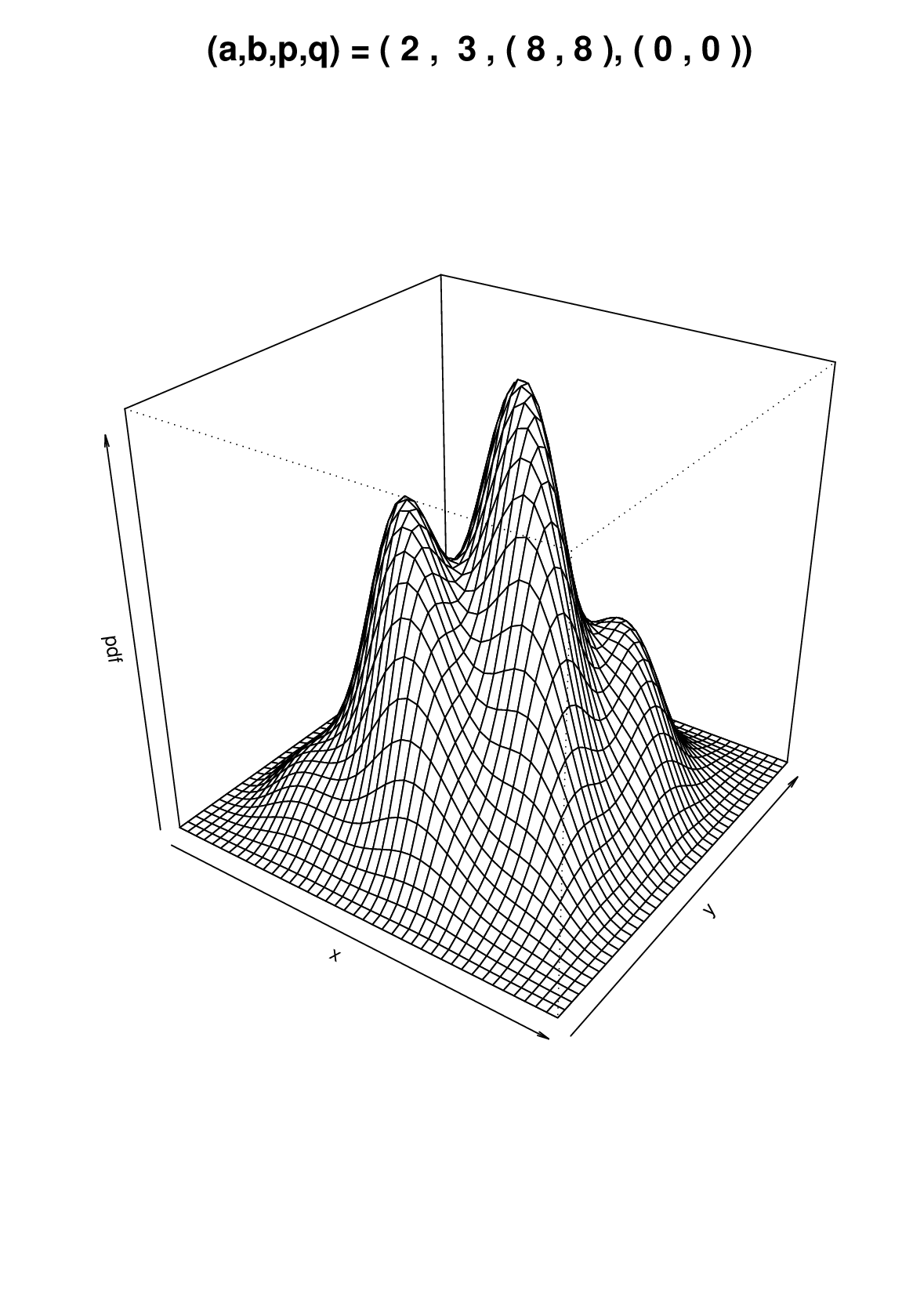}
  } 
\vspace{1ex} 
\centerline{
  \includegraphics[width=0.50\hsize,height=68.5mm]{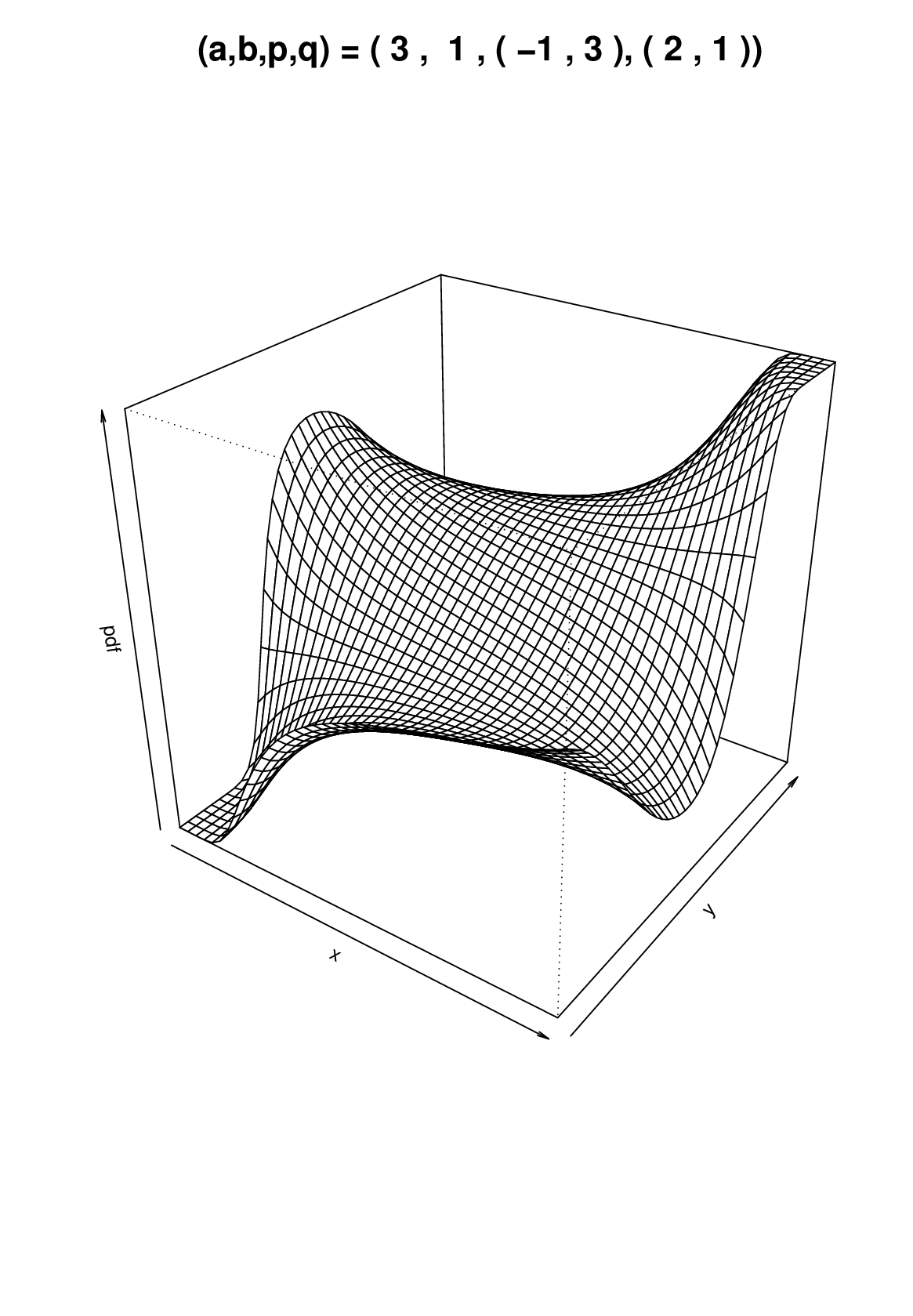}
  \,
  \includegraphics[width=0.50\hsize,height=68.5mm]{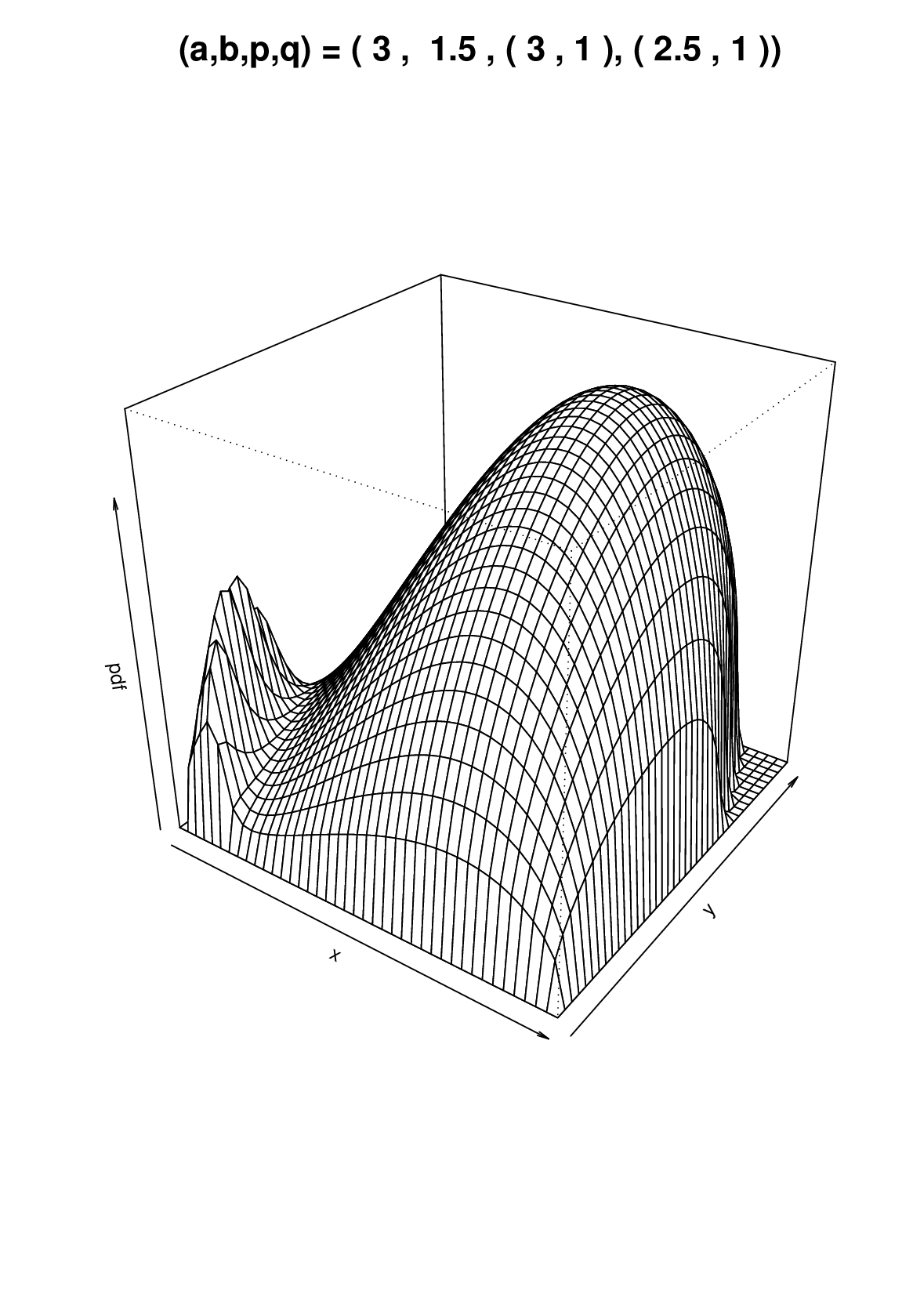}
 } 
\vspace{1ex} 
\centerline{
  \includegraphics[width=0.50\hsize,height=68.5mm]{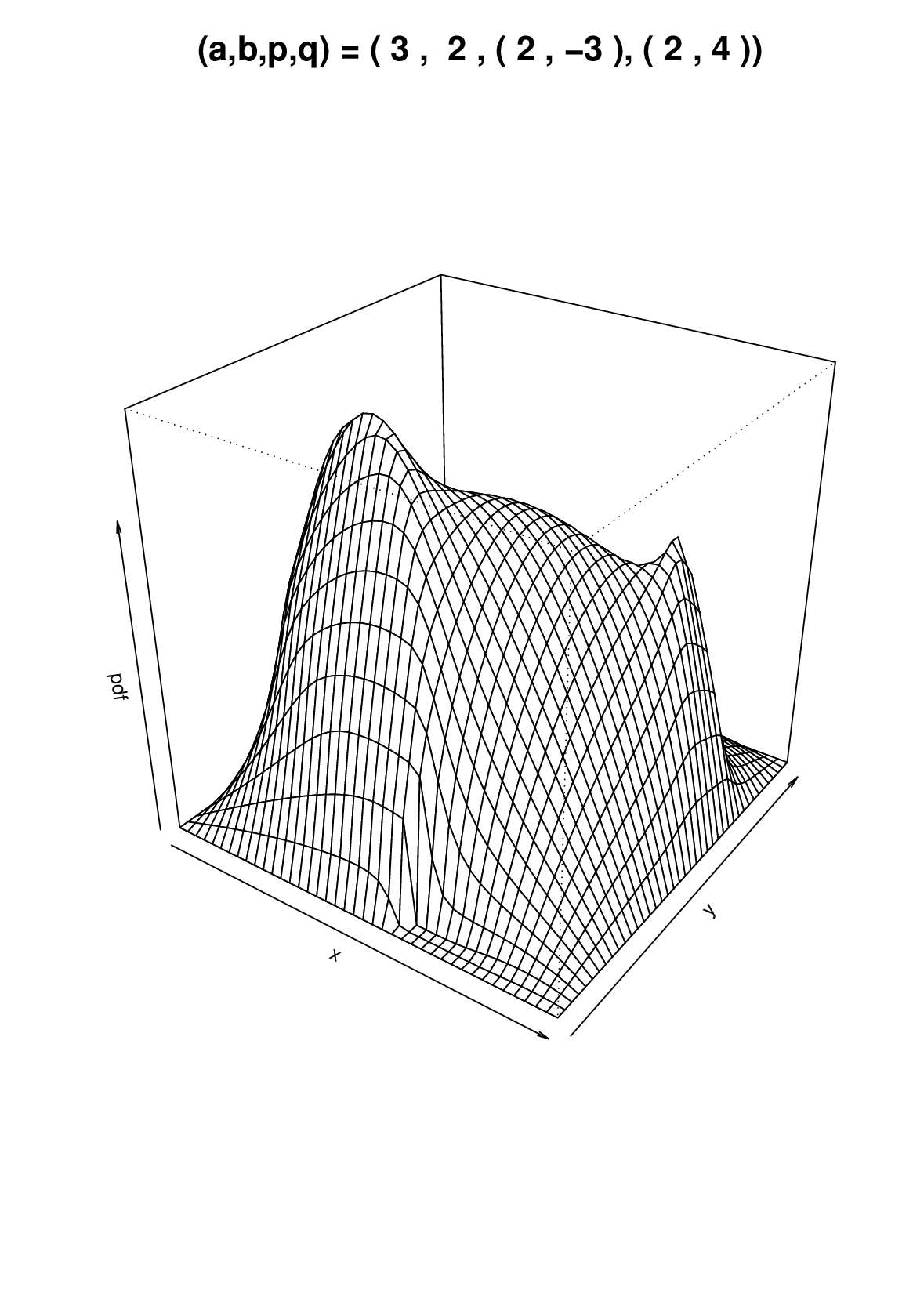}
  \,
  \includegraphics[width=0.50\hsize,height=68.5mm]{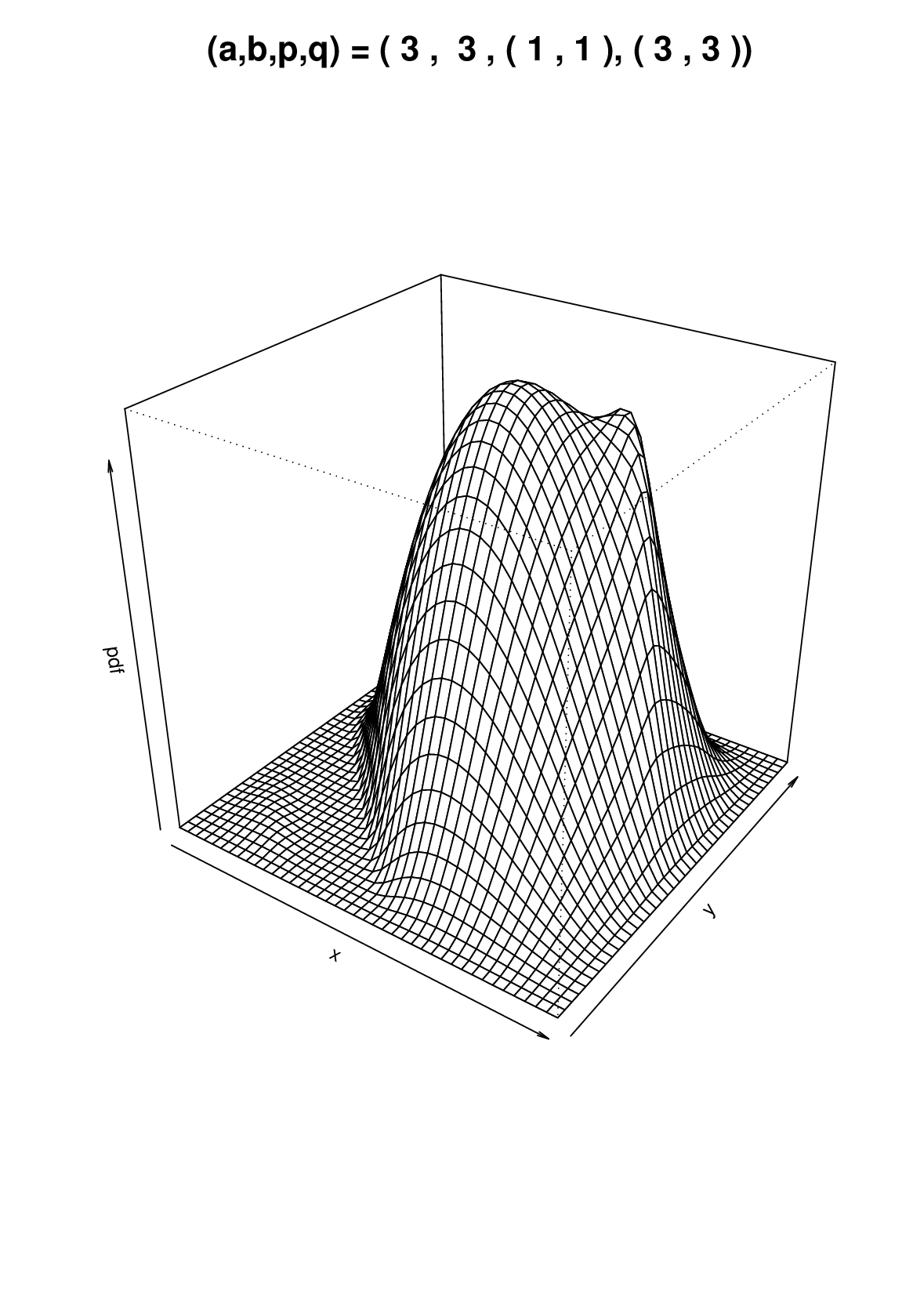}
 } 
\caption{\small Examples of perturbed symmetric Beta densities.
  The set of parameters $(a, b, p_1, p_2, q_1, q_2)$ is shown at
  the top of each plot}
 \label{fig:beta}
\end{figure}

We have generated several plots of the above type of density,
obtaining an extremely rich set of surfaces, as indicated by the small
collection of such densities given in Figure~\ref{fig:beta}.
Additional regulation of the shape could be achieved, by inserting
parameters in the logistic function $G(x)$, although it is doubtful
that one would need the latter level of additional flexibility.  The
plots indicate that the effect of perturbing $f$ via (\ref{f:pdf}) is
far more complex than the effect introduced, say, by the skewing
factor of the normal density in (\ref{f:sn}).  Clearly, the purpose of
Figure~\ref{fig:beta} is purely illustrative, and it is not suggested
to use the above class of density functions in practice without
further investigation.

For a random variable with density (\ref{f:pdf}), the stochastic
representation given by Azzalini \& Capitanio (1999, p.\,599) for
a slightly different case is still valid. In fact, the conditions
required there for its validity are actually those of
Proposition~\ref{p:df-se}. Specifically, if $Y$ has density
function $f$ and $X$ is an independent variable with distribution 
function $G$, then
\begin{equation}  \label{f:genesis}
  Z = \cases{ Y & if $X<w(Y)$ \cr
             -Y  & if $X>w(Y)$
           }
\end{equation}
has density function (\ref{f:pdf}). Clearly, this provides an
algorithm for generating $Z$ and it will also turn out to be
useful for theoretical purposes.

It can be shown that the conditioning method for generating skew
normal random variables from  (\ref{f:sr-condit}) is a special
case of (\ref{f:genesis}). In fact, from consideration of the
residual part of $U_0$ after removing the regression on $U$,
define the variable
\begin{equation}  \label{f:Xtilde}
  \tilde{X} = - \recradice{1-\delta\T\bar\Omega\inv\delta}
             (U_0- \delta\T\bar\Omega\inv U) \sim  N(0,1),
\end{equation}
independent of $U$. After substituting symbols, the condition
$\tilde{X}<\alpha\T U$ of the top branch of (\ref{f:genesis}) is
equivalent to $U_0>0$, if $\alpha$ is given by (\ref{f:alpha});
hence it generates a $\SN_d(0,\bar\Omega, \alpha)$ variable if we
set $Z=U$. The condition of the lower branch is equivalent to
$-\tilde{X} < \alpha\T(-U)$ leading to a $\SN_d(0,\bar\Omega,
-\alpha)$ variable if we set $Z=U$, hence to a
$\SN_d(0,\bar\Omega, \alpha)$ variable if we set $Z=-U$.

Similarly, the stochastic representation of a variate with density
(\ref{f:esn}) via $(U|U_0+\tau>0)$ could be reformulated in terms
of the condition $\tilde{X}<\alpha_0+\alpha\T U$. In general, the
existence of a similar correspondence would be unclear if the
assumption of normality in (\ref{f:sr-condit}) was replaced by
some other distributional assumption. Luckily, a suitable
transformation analogous to (\ref{f:Xtilde}) can be obtained in a
few important special cases to be discussed in Section~\ref{s:se}.

It is immediate that, if $f$ is an elliptical density, $G$
corresponds to a  distribution symmetric about 0 and
$w(y)=\alpha\T y$ for some $\alpha\in\Real^d$, then the conditions
required by Proposition~\ref{p:df-se} are fulfilled. We then
obtain the family of densities produced by Corollary~2 of Azzalini
\& Capitanio (1999).

\begin{proposition}  \label{th:transf}
Denote by $Y$ and $Z$  two $d$-dimensional  random variates having
density function $f$ and (\ref{f:pdf}), respectively, satisfying
the conditions of Proposition~\ref{p:df-se}. If $t(\cdot)$ is a
function from $\Real^d$ to some Euclidean space, such that
$t(-y)=t(y)$ for all $y\in\Real^d$, then
\[  t(Y) \equald t(Z)\,.  \]
\end{proposition}
\emph{Proof}. This is immediate from representation
(\ref{f:genesis}).

A key example of the above result is obtained when $t(y)$
represents the distance from the origin. Since any choice of
$t(\cdot)$ must satisfy the symmetry condition $t(y)=t(-y)$, then
the probability distribution of the distance of a random point
from the origin is the same for $Y$ and for $Z$. In particular we
can write $Y\T B Y \equald Z\T B Z$ for any positive definite
matrix $B$.  A result similar to Proposition~\ref{th:transf}
for the case when $f$ is an elliptical distribution has been
given by Genton \& Loperfido (2002).

A related set of applications of Proposition~\ref{th:transf} is
offered by various results on quadratic forms of skew normal
variates, all of which lead to the conclusion that known
distributional results for normal variates still hold if the
variates are of skew normal type. This set of results includes
Proposition 7, 8 and 9 of Azzalini \& Capitanio (1999) and
Proposition~1, 2 and 6 (parts 1 and 3) of Loperfido (2001). For
these conclusions, one must consider functions $t(\cdot)$ in
Proposition~\ref{th:transf} taking on values in an appropriate
Euclidean space, for instance $\Real^+\times\Real^+$ if the
independence of two quadratic forms is under consideration. Notice
that Propositions~8 and 9 of Azzalini \& Capitanio (1999) have
added  conditions on the $\alpha$ parameter, but these are not
necessary. There is no conflict with the present conclusions since
in their Proposition~8 this extra condition is part of a
sufficiency requirement, and their Proposition~9 (a Fisher-Cochran
type of theorem) was stated in a more restricted form than 
actually possible.

We conclude this section with a discussion on possible
generalisations of Proposition~\ref{p:df-se}. A very general form
of density resembling (\ref{f:pdf}) is along the following lines.
  Denote by $X=(X_1,\dots, X_m)\T$ an $m$-dimensional random variable
  with distribution function $G$, by $Y$ an independent  $d$-dimensional
  random variable  with density function $f$, and by $w_1(y),\dots,w_m(y)$
  a set of   functions from $\Real^d$ to $\Real$.
  For the moment, we remove any assumptions on $f$, $G$ and the
  $w_i$'s; there is no loss of generality in assuming $w_i(0)=0$,
  since  otherwise $w_i(0)$ could be absorbed into the  $b_i$'s to 
  be introduced in a moment.
Then
\begin{equation}  \label{f:pdf-general}
           p\inv\: G\{w_1(y)+b_1, \dots,w_m(y)+b_m\}\, f(y)
\end{equation}
is a density function for any choice of the real numbers
$b_1,\dots,b_m$, if
\[
   p= \pr{X_1-w_1(Y) \leq b_1,\dots, X_m-w_m(Y)\leq b_m}.
\]
The statement follows immediately from the fact that
\begin{eqnarray*}
   p &=& \E[Y]{\pr{X_1-w_1(y) \leq b_1,\dots, X_m-w_m(y)\leq b_m |Y=y}}
\\
   &=&\int_{\Real^d} G\{w_1(y)+b_1, \dots,w_m(y)+b_m\}\, f(y)\d{y}.
\end{eqnarray*}

Clearly, the difficulty is in computing the normalising constant
$p$. This task is amenable when $X$ and $Y$ are multivariate
normal variables. A rather simple special case of
(\ref{f:pdf-general}) is given by (\ref{f:esn}) where $G$ is the
scalar normal distribution function, and $f$ is
$\phi_d(x;\Omega)$. An instance of density (\ref{f:pdf-general})
with multivariate $G$ is given by Sahu \emph{et al.}  (2001); in
their case, $f$ is the $d$-di\-men\-sional normal density, $G$ is
the $d$-dimensional normal distribution function, the $w_j$'s are
$d$ linear combinations of $y$ and all $b_j$'s are 0. The
multivariate distribution sketched by Azzalini (1985, section~4)
and the multiple constraint model outlined by Arnold and Beaver
(2000a, section~6) has a $G$ which is the product of $m$ ($m\ge
1$) terms of type $\Phi(\alpha_i y_i)$ or $\Phi(\alpha\T_i y +
b_i)$, respectively. The  `general multivariate skew normal
distribution' mentioned by Gupta, Gonzáles-Farías and
Domínguez-Molina (2001, section~5) is even more general since they
adopt a $G$ which is the the $m$-dimensional normal distribution
function.

When $f$ or $G$ or both, in (\ref{f:pdf-general}), are not of
Gaussian type, evaluation of $p$ is generally much more
problematic. Some form of restrictions must however be imposed,
not only to make the problem tractable but also because it has
little meaning to consider (\ref{f:pdf-general}) in its full
generality which is so broad as to lose nearly any structure. A
reasonable setting is as follows: suppose  that $f$ and $G$ are
both centrally symmetric and $w_i(-y)=-w_i(y)$ for all
$y\in\Real^d$. Then, by using  essentially the same argument as in
the proof of Proposition~\ref{p:df-se}, one concludes that
$W=(W_1,\dots,W_m) = (w_1(Y),\dots,w_m(Y))$ is centrally
symmetric; therefore so is $V=(X_1-W_1, \dots, X_m-W_m)$, by using
the properties of centrally symmetric functions. A tractable
instance of this setting is offered by the skew Cauchy
distribution and its variants discussed by  Arnold and Beaver
(2000b), using a univariate $G$. Exploration of other cases along
the direction sketched above seems very interesting  but far
beyond the scope of the present paper.


\section{Skew elliptical densities}            \label{s:se}

This section focuses on an important subclass of  (\ref{f:pdf})
with the component $f$ of elliptical form, aiming at three main
goals.  The first is to prove that the two forms of skew
elliptical densities introduced by Azzalini \& Capitanio (1999,
p.\,599) and by Branco \& Dey (2001) are closely connected. The
second goal is to show that the relationships among the  three
forms of stochastic representation of a skew normal variate
recalled in Section~\ref{s:prelim}  carry over to skew elliptical
variates. Furthermore, an analogue of stochastic representation
(\ref{f:ru}) for elliptical variates is obtained for skew
elliptical ones.


\subsection{Skew elliptical densities by conditioning}
\label{s:cond-meth}

For simplicity of presentation, we shall work with correlation
matrices, and location parameter 0. For the rest of this section,
$U^*$ denotes a $(d+1)$-dimensional variate partitioned into a
scalar component $U_0$ and a $d$-dimensional vector $U$.

Branco \& Dey (2001) have introduced a class of skew elliptical
distributions generated by applying to a $(d+1)$-dimensional
elliptical variate the same conditioning method described in
Section~\ref{s:prelim} in connection with the SN distribution. The
following proposition recalls their key statement, up to some
inessential changes of notation.

\begin{proposition}
Consider the random vector ${U^*} \sim
\Ell_{d+1}(0,\Omega^*,\gen)$ where $\Omega^*$ is defined in
(\ref{f:sr-condit}). Then the probability density function of
$Z=(U|U_0>0)$ is
\begin{equation}            \label{f:d-se}
     2 f_{U}(z;\bar\Omega)\:\int_{-\infty}^{\alpha\T z}
                 c_1 \, \gen^{Q_z}(y^2)\d y
\end{equation}
where
\begin{equation}  \label{f:Qz}
                 Q_z = z\T \bar\Omega\inv z,
\end{equation}
the vector $\alpha$ is defined in (\ref{f:alpha}), $f_{U}$ is the
density of $U$, $\gen^{Q_z}(\cdot)$ is the density generator
of $(U_0|U=z)$ and $c_1$ is the associated normalizing constant.
\end{proposition}
For later use, note that an alternative expression for
(\ref{f:d-se}) is
\begin{equation}    \label{f:d-se2}
  2\:\int_0^{\infty}
           c_{d+1}\gen\left({u^*}\T (\Omega^*)\inv u^*\right)\:
           |\Omega^*|^{-1/2} \d u_0\,.
\end{equation}

On defining $F^{Q_z}(\cdot)$ to be the distribution function
corresponding to the density generator  $\gen^{Q_z}(\cdot)$, the
above result lead Branco \& Dey (2001) to re-write (\ref{f:d-se})
in the form
\begin{equation}    \label{f:d-se3}
  2\,f_{U}(z;\bar\Omega)\,F^{Q_z}(\alpha\T z)
\end{equation}
where the distribution function $F^{Q_z}$ is actually varying at
each selected point $z$.  This expression appears to be different
from  (\ref{f:pdf}) where a fixed distribution function $F$ is
involved.

However, when the quantity $Q_z$ can be removed from the argument
of the integral in (\ref{f:d-se}) by means of a suitable change in
variable, the resulting density function will become
\begin{equation}   \label{f:pdf-z}
        2\: f_{U}(z;\bar\Omega)\:F\{w(z)\}
\end{equation}
where $F$ is a univariate distribution function and $w$ is such
that $w(z)=h(\alpha\T z,z\T \bar\Omega\inv z)$ for some function
$h$ from $\Real\times\Real^+$ to $\Real$.  It is easy to show
that the property  $w(-z)=-w(z)$ must hold; hence
 (\ref{f:pdf-z}) is of type (\ref{f:pdf}).

It is difficult to state general conditions under which a density
of type (\ref{f:d-se3}) can actually be transformed into one of
form (\ref{f:pdf-z}), but special cases where this is indeed
feasible do exist.  We shall now examine in detail two important
cases of this form, namely when $U^*$ has either a $\PVII_{d+1}$
or a $\PII_{d+1}$ distribution, which are among those considered
by Branco \& Dey (2001).

\begin{proposition}  \label{p:spvii}
If the random vector $U^*$ has a $\PVII_{d+1}(0,\Omega^*,M,\nu)$
distribution, then the probability density function of
$Z=(U|U_0>0)$ is
\begin{equation}    \label{f:pdf-spvii}
    2\: f_{U}(z;\bar\Omega)\:
        F_1\left(\alpha\T z\recradice{\nu+ Q_z};M,1\right),\qquad
    z\in\Real^d,
\end{equation}
where $Q_z$ is given by (\ref{f:Qz}), $f_{U}$ is the density of a
$\PVII_d(0,\bar\Omega,M-1/2,\nu)$ and $F_1(\cdot;M,1)$ is the
cumulative probability function of a $\PVII_1(0,1,M,1)$.
\end{proposition}
\emph{Proof}. Using results in Fang, Kotz and Ng (1990,
pp.\,82--83), we have
\[
c_1\gen^{Q_z}(y^2) = \dfrac{\Gamma(M)}{\pi^{1/2} \Gamma(M-1/2)}
         \recradice{\nu + Q_z}\left(1+\frac{y^2}{\nu + Q_z}\right)^{-M}
\]
and
\[
  f_U(z;\bar \Omega) =\frac{\Gamma(M-1/2)}
                     {|\bar\Omega|^{1/2}(\pi\nu)^{d/2}\Gamma(M-(d+1)/2)}
      \left(1+\frac{Q_z}{\nu}\right)^{-M + 1/2}
\]
i.e. the densities of a $\PVII_1(0,1,M,\nu+Q_z)$ and of a
$\PVII_d(0,\bar\Omega,M-1/2,\nu)$ variate with parameters $M -
1/2$ and $\nu$, respectively. On setting $x = y\recradice{\nu +
Q_z}$, the integral in (\ref{f:d-se}) becomes
\[
 \int_{-\infty}^{\alpha\T z \recradice{\nu + Q_z}}
    \dfrac{\Gamma(M)}{\pi^{1/2} \Gamma(M-1/2)}
         (1+x^2)^{-M}    \d x
\]
which is the distribution function of a $\PVII_1(0,1,M,1)$ variate
evaluated at the point $\alpha\T z \recradice{\nu + Q_z}$. QED

\noindent\emph{Example~1: skew $t$ distribution}. The  relevance
of the $\PVII_d$ class is due to the inclusion of the multivariate
$t$ family as the special case when $M = (d+\nu)/2$. The
corresponding specification of Proposition~\ref{p:spvii} produces
then a form of multivariate skew $t$ density. Since
Section~\ref{s:skew-t} will  be entirely dedicated to this
distribution, we defer detailed discussion until then.

\begin{proposition}  \label{p:spii}
If the $(d+1)$-dimensional elliptical random vector $U^*$ has a
$\PII_{d+1}(0,\Omega^*,\nu)$ distribution, then the probability
density function of $Z=(U|U_0>0)$ is
\begin{equation} \label{f:pdf-spii}
   2 f_{U}(z;\bar\Omega)\,
     F_1\left(\alpha\T z\,\recradice{1-Q_z};\nu \right),
   \qquad z\in(-1,1)^d,
\end{equation}
where $Q_z$ is given by (\ref{f:Qz}), $f_{U}$ is the density of
a $\PII_d(0,\bar\Omega,\nu+1/2)$ variate, and $F_1(\cdot;\nu)$ is the
distribution function of a $\PII_1(0,1,\nu)$.
\end{proposition}
\emph{Proof}. Identical to that of Proposition  \ref{p:spvii},
considering  the densities of marginal and conditional
distributions of $\PII$, as defined in Fang, Kotz and Ng, (1990,
pp.\,89-91).

The absence of $Q_z$ in the conditional density characterizes the
multivariate normal distribution among the members of the
elliptical family. This fact can be used to produce an analogous
characterization of the skew normal distribution within the skew
elliptical family.

\begin{proposition}
The function $w$ in (\ref{f:pdf-z}) is such that $w(z) = \alpha\T
z$ if and only if~$U^*$ is Gaussian, i.e. $Z$ is skew normal.
\end{proposition}
\emph{Proof}. The density of $(U|U_0 = z)$ does not depend on
$Q_z$ if and only if $U^*$ is Gaussian; see Theorem 4.12 of Fang
\emph{et al.} (1990). In this case, the integral in (\ref{f:d-se})
becomes $\Phi(\alpha\T z)$,
so that $Z \sim \SN_d(0,\bar\Omega,\alpha)$.~QED

A number of parallels between the skew normal distribution and
other types of skew elliptical distributions have already been
shown. The next result allows us to construct a random variable
$\tilde{X}$ playing a role analogous to the one in (\ref{f:Xtilde})
for the skew version of a $\PVII_d$ and $\PII_d$ distribution,
respectively.

\begin{proposition} \label{p:e-Xtilde}
Let $U^* \sim \PVII_{d+1}(0,\Omega^*,M,\nu)$. Then
\[
  \tilde{X} =
     -\recradice{1-\delta \bar\Omega\inv \delta}(U_0 - \delta\T
     \bar\Omega\inv U)
     \recradice{\nu +U\T \bar\Omega\inv U}\sim \PVII_1(0,1,M,1),
\]
independent of $U$. If $U^* \sim \PII_{d+1}(0,\Omega^*,\nu)$ then
\[
  \tilde{X} =
     -\recradice{1-\delta \bar\Omega\inv \delta}(U_0 - \delta\T
     \bar\Omega\inv U)
     \recradice{1 - U\T \bar\Omega\inv U} \sim \PII_1(0,1,\nu),
\]
independent of $U$.
\end{proposition}
\emph{Proof}. By direct calculation.

Therefore,   we can set
\[
   Z = \cases{U  &  if $\tilde{X}< w(U)$, \cr
               -U & if $\tilde{X}> w(U)$,}
\]
where $w(z)$ is the transformation of $z$ used in the argument of
$F_1$ in (\ref{f:pdf-spvii}) and (\ref{f:pdf-spii}), respectively;
it is intended that  the appropriate distribution of $U^*$ and
transformation $\tilde{X}$ have been selected. This formula
establishes a method of type (\ref{f:genesis}) to generate a  skew
$\PVII_d$ and skew $\PII_d$ variate, respectively.


The connections between the proposal of Azzalini \& Capitanio
(1999) and the one of Branco \& Dey (2001) can be summarised as
follows. The conditioning argument which is one of the mechanisms
to generate the skew normal distribution from the normal one can
be adopted to generate a form of skew elliptical distributions
from the elliptical ones, leading to (\ref{f:d-se3}), or some
similar form as obtained by Branco \& Dey.  This type of
expression can, at least in some important special cases, be
transformed into one where the skewing factor of $f$ is a fixed
distribution function, as shown by (\ref{f:pdf-spvii}) and
(\ref{f:pdf-spii}).  These expressions are of type (\ref{f:pdf}),
which is essentially the form of Azzalini \& Capitanio.  The
natural question is whether all densities of type (\ref{f:d-se3})
can be re-written in the form (\ref{f:pdf}), but we have been
unable to prove this fact in general. Notice that the converse
inclusion is not true, that is, not all densities of type
(\ref{f:pdf}) can be written in the form (\ref{f:d-se3}), unless
additional restrictions are imposed on the components of
(\ref{f:pdf}), besides the obvious condition that $f$ is
elliptical.

The next result concerns a stochastic representation of type
(\ref{f:ru}) for distributions of type (\ref{f:pdf}) when the
density $f$ is elliptical. For example, this representation is
valid for the skew elliptical densities defined in Azzalini \&
Capitanio (1999, p.\,599) and for the skew versions of $\PVII_d$
and $\PII_d$ examined earlier.

\begin{proposition}  \label{p:stoch-rep}
If $Z$ has a density of type (\ref{f:pdf}), where $f$ is the
density of $U \sim \Ell_d(\xi,\bar\Omega,\gen)$, then $Z$ admits
the stochastic representation
\begin{equation}  \label{f:polar}
Z = \xi + R L\T S'
\end{equation}
where $\bar\Omega = L\T L$, $R>0$ has the same distribution as the
radius of the stochastic representation (\ref{f:ru}) of $U$, and
$S'$ has a non-uniform distribution on the unit sphere of
$\Real^d$. Specifically, using spherical coordinates, the density
of $S'$ is equal to
\[
    \frac{\Gamma(d/2)}{\pi^{d/2}}
       \prod_{k=1}^{d-2}(\sin\theta_k)^{d-k-1}
               \pr{X\leq w_L^*(\theta_1,\ldots,\theta_{d-1},R)},
\]
where $w_L^*(\cdot)$ is a function from $\Real^d$ to $\Real$
defined in Appendix A, and $X$ is an independent random variable
having distribution function $G$. Furthermore, the conditional
distribution of $S'$ given $R=r$ is of type (\ref{f:pdf}), with
density
\[
  \dfrac{\Gamma(d/2)}{\pi^{d/2}}\prod_{k=1}^{d-2}(\sin\theta_k)^{d-k-
1}
  G\{w_L^*(\theta_1,\ldots,\theta_{d-1},r)\}.
\]
\end{proposition}
\emph{Proof}. In Appendix A.

\noindent \emph{Example~2: Stochastic representation
(\ref{f:polar}) for skew normal distribution}. If $Z \sim
\SN_d(\xi,\bar \Omega,\alpha)$, then by applying Proposition
\ref{p:stoch-rep} we obtain $R^2 \sim \chi_d^2$ and the following
spherical coordinates representation of the marginal distribution
of $S'$:
\begin{eqnarray*}
   f_{\theta}(\theta) & = & 2\,\frac{\Gamma(d/2)}{2\pi^{d/2}}
                 \prod_{k=1}^{d-2}(\sin\theta_k)^{d-k-1}\\
             &   & \qquad \pr{X\leq
                  R\,(\alpha_1^*\cos\theta_1+\alpha_2^*\sin\theta_1\cos\theta_2+\cdots+
                  \alpha_d^*\sin\theta_1\cdots \sin\theta_{d-1})},
\end{eqnarray*}
where $\theta=(\theta_1,\theta_2,\ldots,\theta_{d-1})\T$,
$\alpha^* = L\,\alpha$ and $X \sim \N(0,1)$ is independent of $R$.
Finally, noticing that $d^{1/2}X \,R^{-1}$ has a $t$ distribution
with $d$ degrees of freedom, we have
\begin{eqnarray*}
   f_{\theta}(\theta) & = & 2 \,\frac{\Gamma(d/2)}{2\pi^{d/2}}
                   \prod_{k=1}^{d-2}(\sin\theta_k)^{d-k-1}\\
   &   & \qquad T_1\{d^{1/2}(\alpha_1^*\cos\theta_1+\alpha_2^*\sin\theta_1\cos\theta_2+
              \ldots + \alpha_d^*\sin\theta_1\cdots \sin\theta_{d-1});d\}
\end{eqnarray*}
where $T_1(\cdot;d)$ is the distribution function of a scalar $t$
distribution with $d$ degrees of freedom.


\subsection{Skew elliptical densities by transformation method}

The next result shows how the class of skew elliptical
distributions mirrors another property of the skew normal
distribution. In fact the class of skew elliptical densities
obtained via the conditioning method is equivalent to the one
obtained by applying the transformation method recalled in Section
\ref{s:prelim}.

\begin{proposition}  \label{p:tr-meth}
Consider the random vector $(U_0, U) \sim
\Ell_{d+1}(0,\Psi^*,\gen)$ where $\Psi^*$ is as in
(\ref{f:Omega^*}), and define
\begin{equation}    \label{f:sr-transf2}
    Z_j = \delta_j\: |U_0| + \radice{1-\delta_j^2}\: U_j,
   \qquad j=1,\dots,d,
\end{equation}
where $-1<\delta_j<1$. Then the density of $(Z_1,\ldots,Z_d)$ is
of type (\ref{f:d-se}), where
\begin{eqnarray*}
  \lambda_i & = & \delta_i \recradice{1-\delta_i^2}, \qquad(i=1,\dots,d
),\\
  \Delta & = & \diag\{\recradice{1+\lambda_1^2},\dots,
              \recradice{1+\lambda_d^2}\},\\
  \Omega & = & \Delta(\Psi+\lambda \lambda\T)\Delta, \\
  \alpha & = & \recradice{1+\lambda\T \Psi \lambda}\Delta\inv
                                                \Psi\inv \lambda\,.
\end{eqnarray*}
\end{proposition}
\emph{Proof}. First note that the joint density function of
$|U_0|$ and $U$ takes the form $2c_{d+1}\gen(\cdot)$. Denote by
$B$ the $(d+1)\times(d+1)$ matrix implicitly defined by
(\ref{f:sr-transf2}) such that $(Z_0,Z_1,\ldots, Z_d)\T= B
(|U_0|,U\T)\T$, and apply the usual formulae for linear
transforms. Then the density function of $(Z_1,\dots,Z_d)$ turns
out to be
\[
  2\int_0^{\infty} c_{d+1}
     \gen\left((z_0,z\T) A\inv (z_0,z\T)\T\right)\,|A|^{-1/2} \d x_0
\]
where $A=B\Psi^* B\T$ is a correlation matrix. Taking into account
expression (\ref{f:d-se2}) the result follows. QED

An immediate consequence of the transformation method is a further
generating method for the bivariate case. Again, this reproduces
for the skew elliptical family a generation method known to hold
for the skew normal distributions.


\begin{proposition}
If $(U_0,U) \sim \Ell_2(0,\Omega^*,\gen)$, the class generated by
$Z = \max(U_0,U)$ is equal to the class generated by  the
transformation method of Proposition~\ref{p:tr-meth} with $d=2$.
\end{proposition}
\emph{Proof}. First notice that $\max(U_0,U) = \half |U-U_0|+
\half (U+U_0)$. As the joint distribution of
$(U-U_0)\recradice{2-2\rho}$ and $(U+U_0)\recradice{2+2\rho}$ is
$\Ell_2(0,I,\gen)$, where $\rho$ denotes the off-diagonal elements
of $\Omega^*$, the result follows by direct application of
Proposition \ref{p:tr-meth} on imposing
$\delta=\radice{\half(1-\rho)}$. QED


\section{A skew $t$ distribution}    \label{s:skew-t}

For the rest of the paper we shall focus on the development of an
asymmetric version of the multivariate Student's $t$ distribution,
already sketched in Section~\ref{s:cond-meth}. The purpose of the
present section is to provide additional support for its
definition and to examine more closely its properties. Connected
inferential aspects will be discussed in the subsequent section.

\subsection{Definition and density}

The usual construction of the $t$ distribution is via the ratio of
a normal variate and an appropriate transformation of a
chi-square.  If one wants to introduce an asymmetric variant of
the $t$ distribution, a quite natural option is to replace the
normal variate above by a skew normal one.

A preliminary result on Gamma variates is required. We shall say
that a positive random variable is distributed as
$\rvGamma(\psi,\lambda)$ if its density at $x \:(x>0)$ is
\[
    \frac{\lambda^\psi}{\Gamma(\psi)}\,  x^{\psi-1}\,\exp(-\lambda\,x).
\]

\begin{lemma} \label{th:gamma}
If $V \sim \rvGamma(\psi,\lambda)$, then for any $a,\,b\in\Real$
\[
    \E{\Phi(a\sqrt{V}+b)} =  \pr{T\leq a\sqrt{\psi/\lambda}}
\]
where $T$ denotes a non-central $t$ variate with $2\psi$ degrees
of freedom and non-centrality parameter $-b$.
\end{lemma}
\emph{Proof.\quad} Let $U\sim N(0,1)$; then
\begin{eqnarray*}
  \E{\Phi(a\sqrt{V}+b)}
     &=& \E[V]{\pr{U\leq a\sqrt{v}+b|V=v}} \\
     &=& \E[V]{\pr{(U-b)/\radice{v\lambda/\psi}
                \leq a\radice{\psi/\lambda}|V=v}}\\
     &=& \pr{T'\leq a\radice{\psi/\lambda} }
\end{eqnarray*}
where $T= (U-b)/\radice{V\lambda/\psi} $ has the quoted $t$
distribution. ~QED

As anticipated earlier, we define the skew $t$ distribution as the
one corresponding to the transformation
\begin{equation} \label{f:Y}
               Y = \xi + V^{-1/2}\,Z
\end{equation}
where $Z$ has density function (\ref{f:sn}) with $\xi=0$, and
$V\sim\chi^2_\nu/\nu$, independent of $Z$. An equivalent
interpretation of  $Y$ is to regard it as a scale mixture of SN
variates, with mixing scale factor $V^{-1/2}$. Application of the
above lemma to a $\rvGamma(\half\nu,\half\nu)$ variate and some
simple algebra lead to the  density of $Y$, which is
\begin{equation} \label{f:pdf-t}
   f_Y(y) = 2  \: t_d(y;\nu)\:
            T_1\left( \alpha\T\omega\inv (y-\xi)\radice{\frac{\nu+d}%
                      {Q_y+\nu}};  \nu+d
              \right)
\end{equation}
where  $\omega$ is defined at the beginning of Section~\ref{s:prelim},
\begin{eqnarray*}
   Q_y &=& (y-\xi)\T\Omega\inv (y-\xi)\,, \\
   t_d(y;\nu) &=& \frac{1}{|\Omega|^{1/2}}\: g_d(Q_y;\nu)
             \:=\:  \frac{\Gamma((\nu+d)/2)}
                  {|\Omega|^{1/2}\,(\pi\nu)^{d/2}\,\Gamma(\nu/2)}
                  (1+Q_y/\nu)^{-(\nu+d)/2}
\end{eqnarray*}
is the density function of a $d$-dimensional $t$ variate with
$\nu$ degrees of freedom, and $T_1(x;\nu+d)$ denotes the scalar
$t$ distribution function with $\nu+d$ degrees of freedom. We
shall call distribution (\ref{f:pdf-t}) skew $t$, and write
\begin{equation} \label{f:St}
         Y \sim \St_d(\xi,\Omega,\alpha,\nu)\,.
\end{equation}

It is easy to check that density (\ref{f:pdf-t}) coincides with the
one sketched in Section~\ref{s:cond-meth} using
Proposition~\ref{p:spvii}, which is of type (\ref{f:pdf}).  Moreover,
for the reasons explained in that section, (\ref{f:pdf-t}) coincides
in turn with the skew $t$ distribution of Branco \& Dey (2001),
although this equality is not visible from their derivation because
they did not provide the above closed-form expression of the density.

Therefore, we have seen that a number of different ways to define
a skew $t$ distribution all lead to the same density
(\ref{f:pdf-t}). While additional proposals to introduce a form of
a skew $t$ density are possible, this one has the advantage of
arising from various generating criteria, which in turn are linked
to other portions of literature.

A reviewer of this paper has remarked that, if we set $d=1$,
density (\ref{f:pdf-t}) does not reduce to the form
$2\,t_1(y;\nu)\,T_1(\alpha y;\nu)$, which seems to be the `most
natural' univariate form of skew $t$ density generated by Lemma~1
of Azzalini (1985), a forerunner of Proposition~\ref{p:df-se}.
While the latter density has the appeal of a slightly simpler
mathematical expression, the arguments indicated in the previous
paragraph lead us to prefer  (\ref{f:pdf-t}). In fact, one could
reverse the reasoning, and claim that  Lemma~1 of Azzalini (1985)
`should' had been stated in the form of Proposition~\ref{p:df-se}
for $d=1$; in other words, there is no reason to restrict $w(y)$
to the linear form $\alpha y$, especially outside the normal case.

Alternative proposals of univariate skew $t$ distributions have been
made by Fernández \& Steel (1998), constructed similarly to the
so-called two-piece normal density, and by Jones (2001), developed by
Jones \& Faddy (2002), which is based on a suitable transformation of
a beta density. A multivariate form of skew $t$ distribution has been
proposed by Jones (2002) but the associated inferential aspects have
not been discussed. The alternative form of multivariate skew $t$
distribution considered by Sahu \emph{et al.} (2001) concides with
(\ref{f:pdf-t}) in the case $d=1$; for general $d$, their density
involves the multivariate $t$ distribution function. The density
examined in this paper  allows a relatively simple mathematical
treatment, and it is more naturally linked to the skew normal
distribution, via mechanisms already mentioned. As a consequence,
the distribution enjoys various useful formal properties, which will
be examined in the remaining part of this section.


\subsection{Some properties}
\paragraph{Distribution function}

For simplicity of exposition, we obtain the distribution function
of $Y$ in the `standard' case with $\xi=0, \Omega=\bar\Omega$.
Bearing in mind the  representation of $Z$ based on conditioning,
write
\begin{eqnarray*}
  \pr{Y\leq y} &=& \pr{ V^{-1/2}\,Z \leq y}  \\
          &=& \pr{V^{-1/2}\,U \leq y \:|\: U_0>0} \\
          &=& 2\:\pr{V^{-1/2}\pmatrix{ -U_0 \cr U } \leq\pmatrix{0\cr y}}
                  \\
          &=& 2\: \pr{T' \leq \pmatrix{0\cr y}}
\end{eqnarray*}
where $(U_0,U)$ has distribution (\ref{f:sr-condit}), and the
inequality signs are intended componentwise. The last expression
involves the integral of a multivariate $(d+1)$-dimensional $t$
variate $T'$ with dispersion  matrix similar to the one of
(\ref{f:sr-condit}), but with reversed sign of $\delta$.
Algorithms for computing this type of distribution function are
given by Genz \& Bretz (1999).

An alternative expression for the above distribution function is
given by
\[   \pr{Y\leq y}
       = \pr{V^{-1/2}\,U \leq y \:|\: U_0>0}
       = \E[V]{F_Z(y v^{1/2})|V=v},
\]
where $F_Z$ denotes the distribution of $Z$, hence evaluating the
distribution function of $Y$ by suitably averaging the distribution of
$Z$ with respect to the distribution of $V$.  This  expression is
most useful in the case $d=1$ where a practical expression of $F_Z$ is
available; see formula (4) and subsequent remarks of Azzalini (1985).

\paragraph{Moments}
Using the representation (\ref{f:Y}), it is easy to compute the
moments of $Y$.  For algebraic convenience, we assume $\xi=0$
throughout.  If $\E{Y^{(m)}}$ denotes a moment of order $m$, write
\begin{equation} \label{f:mom}
   \E{Y^{(m)}}  = \E{V^{-m/2}} \,\E{Z^{(m)}}
\end{equation}
where  $Z$ has density function (\ref{f:sn}) with $\xi=0$. It is
well-known that
\[
  \E{V^{-m/2}} =
      \frac{(\nu/2)^{m/2}\:\Gamma(\half(\nu-m))}{\Gamma(\half\nu)},
\]
while, for the expressions of \E{Z^{(m)}}, we use results given by
Azzalini \&  Capitanio (1999) and by Genton \emph{et al.} (2001).

First, we apply  (\ref{f:mom}) to the scalar case. On defining
\begin{equation} \label{f:mu}
    \mu =  \delta\:\radice{\nu/\pi}\:
              \frac{\Gamma(\half(\nu-1))}{\Gamma(\half\nu)}, \qquad(\nu>1),
\end{equation}
one obtains, for $\xi=0$,
\begin{eqnarray*}
  \E{Y}   &=&   \omega\: \mu, \\
  \E{Y^2} &=&  \omega^2\:\frac{\nu}{\nu-2}, \\
  \E{Y^3} &=& 
            \omega^3 \:\mu\:(3-\delta^2)\:\frac{\nu}{\nu-3} , \\
  \E{Y^4} &=& \omega^4\:\frac{3\,\nu^2}{(\nu-2)(\nu-4)} ,
\end{eqnarray*}
provided that $\nu$ is larger than the corresponding order of the
moment; the first two of the above expressions have been given by
Branco \& Dey (2001).  After some algebra, the indices of skewness
and kurtosis turn out to be
\begin{eqnarray*}
  \gamma_1 &=&  \mu\,\,
   \left[\frac{\nu(3-\delta^2)}{\nu-3} - \frac{3\,\nu}{\nu-2}+2\,\mu^2\right]\,
   \left[\frac{\nu}{\nu-2}-\mu^2\right]^{-3/2}
                                    \qquad (\mathrm{if~}\nu>3),
\\
  \gamma_2 &=&  \left[
       \frac{3\nu^2}{(\nu-2)(\nu-4)}
      - \frac{4\mu^2\nu(3-\delta^2)}{\nu-3}
      + \frac{6\mu^2\nu}{\nu-2}
      - 3\mu^4\right]\,
   \left[\frac{\nu}{\nu-2}-\mu^2\right]^{-2} -3
                          \qquad (\mathrm{if~}\nu>4).
\end{eqnarray*}
\par
In the multivariate case, we obtain from  (\ref{f:mom}) that
$\E{Y}=\omega\mu$ still holds, provided $\nu>1$ and (\ref{f:mu})
and $\omega$ are intended in vector and matrix form, respectively;
furthermore
\[
  \E{Y\,Y\T} = \frac{\nu}{\nu-2} \: \Omega  \qquad (\mathrm{if~}\nu>2)\,,
\]
leading to
\[  \var{Y}
     = \frac{\nu}{\nu-2}\,\Omega- \omega\mu\mu\T\omega\,.
\]


\paragraph{Linear and quadratic forms} \label{s:LQ}

Consider the affine transformation $a+AY$ where $a\in\Real^m$ and
$A$ is a $m\times d$ constant  matrix of rank $m$. Using
(\ref{f:Y}) we can write
\[  a+AY =  \xi'+ V^{-1/2} AZ \]
where $\xi'=a+A\xi$.  Take into account that
\[ AZ\sim SN_m(0, A\Omega{}A\T,\alpha') \]
on the ground of results given by Azzalini \& Capitanio (1999)
where the explicit expression for $\alpha'$ is given; similar
results, but in a more convenient form, are provided by Capitanio
\emph{et al.} (2003, Appendix A.2). Therefore we obtain
\[
     a+AY \sim \St_m(\xi', A\Omega{}A\T, \alpha',\nu).
\]
In particular for a single component, $Y_r$ say
($r\in\{1,\dots,d\}$), one has
\[
    Y_r \sim\ \St(\xi_r,\omega_{rr}, \alpha_r' ,\nu)
\]
where $\alpha_r'$ is given by (10) of Capitanio \emph{et al.}
(2003).

Similarly, for a quadratic form, $Q= (Y-\xi)\T B (Y-\xi)$, where
$B$ is a symmetric  $d\times d$ matrix, we can write
\[  Q = Z\T B Z / V. \]
For appropriate choices of $B$, the distribution of $Z\T B Z$ is
$\chi^2_{\nu'}$ for some value $\nu'$ of the degrees of freedom.
One such case is (\ref{f:chi2}), where $B=\Omega\inv$. Azzalini \&
Capitanio (1999, Section~3.3) consider more general forms of $B$;
see also Genton \emph{et al.} (2001) for additional results. In
all cases when the $\chi^2$ property holds for $Z$, we can state
immediately
\[  Q/\nu' \sim F(\nu',\nu). \]

This property allows us to produce Healy's-type plots (Healy,
1968) as a diagnostic tool in data fitting, similarly to the
Normal and SN case, just using the Snedecor distribution as the
reference distribution instead of the $\chi^2$. This device will
be illustrated in the subsequent numerical work.


\paragraph{An extended skew $t$ distribution}
If the component $Z$ in (\ref{f:Y}) is taken to have distribution
(\ref{f:esn}) rather than (\ref{f:sn}), we obtain a density which
parallels the role of (\ref{f:esn}) for skew $t$ densities; this
is now discussed briefly.

By using again Lemma~\ref{th:gamma}, the new density turns out to
be of type (\ref{f:pdf-t}), except that $T_1$ refers now to a $t$
distribution with non-centrality parameter $
-\tau(1-\delta\T\bar\Omega\inv\delta)^{-1/2}$ and the normalizing 
constant 2 is replaced by $1/\Phi(\tau)$.
%
The distribution function is obtained with the same sort of
argument of the case $\tau=0$, namely
\begin{eqnarray*}
  \pr{Y\leq y}
          &=& \pr{V^{-1/2}\,U \leq y \:|\: U_0+\tau>0} \\
          &=& \pr{V^{-1/2}\pmatrix{ -U_0-\tau \cr U } \leq\pmatrix{0\cr y
}}
                 /\Phi(\tau) \\
          &=&  \pr{T'' \leq \pmatrix{0\cr y}}/\Phi(\tau)
\end{eqnarray*}
where now $T''$ refers to a non-central multivariate  $t$;
unfortunately, the latter distribution function is appreciably
harder to compute in practice than the analogous one for the
central case. Moments can be computed again with the aid of
(\ref{f:mom}). Those of  the first and second order are, if
$\xi=0$,
\begin{eqnarray*}
  \E{Y}     &=& \E{V^{-1/2}} \zeta_1(\tau)\omega\delta,
                \qquad(\nu>1),\\
  \E{Y\,Y\T} &=&  \frac{\nu}{\nu-2} \: \left(\Omega +
         [\zeta_2(\tau)+\zeta_1^2(\tau)]
         \:\omega\:\delta\,(\omega\delta)\T\right),
            \qquad (\nu>2),
\end{eqnarray*}
where
\[  \zeta_r(x) =\frac{\d^r}{\d x^r}\zeta_0(x), \qquad(r=1,2,\dots), \]
and $\zeta_0$ is defined by (\ref{f:delta-zeta}).

\section{Statistical aspects of the skew $t$ distribution} \label{s:t-stat}

\subsection{Likelihood inference}

Consider $n$ independent observations satisfying a regression
model of type
\[
    y_i \sim \St_d(\xi_i,\Omega,\alpha,\nu), \quad \xi_i=  \beta\T x_i
\]
for $i=1,\dots,n$; here $x_i$ is a $p-$dimensional vector and
$\beta$ is a $p\times d$ matrix of parameters. Also let
\[  X  = (x_1,x_2,\ldots,x_n)\T  \]
be the $n\times p$ design matrix. Notice that we are effectively
considering a multivariate regression model with  error term of
skew $t$ type. It would be inappropriate to use such a
distribution, and in fact even a regular elliptical distribution,
for the joint modelling of the $n$ observations, since usually
these are supposed to behave independently.

It is convenient to reparametrize the problem by writing
\[  \Omega\inv = A\T \diag(e^{-2\rho}) A = A\T D A,
\quad \mathrm{and} \quad
    \eta=\omega\inv \alpha
\]
where $A$ is an upper triangular $d\times d$ matrix with diagonal
terms equal to 1 and $\rho\in\Real^d$.
The loglikelihood function for the parameter $\theta=(\beta, A,
\rho, \eta, \log\nu)$ is then
\begin{equation}\label{f:logL-t}
  \ell(\theta) = \sum_{i=1}^n \ell_i(\theta)
\end{equation}
where $\ell_i(\theta)$ is the contribution to the loglikelihood
from the $i$-th individual; this term is
\[
   \ell_i(\theta) = \log 2 + \half\log|D| + \log g_d(Q_i;\nu)
                  +\log T_1(t(L_i, Q_i, \nu);\nu+d)
\]
where
\[  u_i=y_i-\beta\T x_i,\qquad
    Q_i = u_i\T \Omega\inv u_i, \qquad
    L_i = \alpha\T \omega\inv u_i, \qquad
    t(L,Q,\nu) = L\radice{\frac{\nu+d}{Q+\nu}}.
\]
Maximisation of this log-likelihood function must be accomplished
numerically. To improve efficiency, the derivatives of
(\ref{f:logL-t}) can be supplied to an optimisation algorithm;
details for computing these derivatives are given in an appendix.

A suite of \textsf{R} routines for evaluating the above
log-likelihood and its derivatives has been developed, and it is
available on the WWW at 
\texttt{http://azzalini.stat.unipd.it/SN}.

In connection with the skew normal distribution, Azzalini (1985)
and Azzalini \& Capitanio (1999) have highlighted  some
problematic aspects of the likelihood function. A key feature is
that the profile log-likelihood function for $\alpha$ always has a
stationarity point at $\alpha=0$, which in turn is connected to
singularity of the information matrix at $\alpha=0$.  These
problematic features were the motivation to introduce an
alternative parametrization which overcomes most if not all of
these problems.

It was a pleasant surprise to find that in the present setting the
behaviour of the log-likelihood function was to be much more
regular, at least for those numerical cases which we have
explored.  A graphical illustration of this statement is given by
Figures~\ref{fig:glass-logL} and \ref{fig:returns-logL} below, 
which show some profile log-likelihood plots. 
These plots refer  to specific datasets, but a similar regularity 
was found with some other datasets which we have considered.

It would be useful to have some theoretical insight on why the
log-likelihood function using the skew $t$ distribution behaves so
differently from the skew normal model, as well as to gather  more
numerical evidence of its behaviour.  However this theme appears
to be a project on its own, and cannot be pursued here.


On another front, Fernández \& Steel (1999) have highlighted
difficulties in regression models when the error term is assumed
to have a $t$ distribution with unspecified degrees of freedom to
be estimated from the data. Specifically, their Theorem~5 states
there are points of the parameter space where the likelihood
function becomes unbounded, if the degrees of freedom are allowed
to span over the whole range $\nu\in(0,\infty)$. To avoid this
effect, one must restrict the range of $\nu$ to the interval
$(\nu_0,\infty)$, where the threshold $\nu_0$ is a function of $X$
and $y$. For instance, in the case of a simple random sample with
no ties in the $y_i$'s, we obtain $\nu_0=d/(n-1)$, which imposes 
a very mild limitation. For the stackloss data example discussed by
Fernández \& Steel (1999) with $d=1$ and $p=3$, the value of
$\nu_0$ is small, 8/13. In addition, they recall some numerical
examples from the literature where poles have been found by
various authors; in all these cases, however, these poles where
found at values of $\nu$ very small, always below 0.30.

Therefore, in practice the difficulties can be circumvented by
avoiding a certain portion of the parameter space which would be
somewhat peculiar anyway. However, the fact that $\nu_0$ depends
on the response variable leads to a procedure which lacks
complete support by the theory of likelihood inference. As
advocated by Fernández \& Steel, a better theoretical
understanding of this sort of model and the associated
log-likelihood properties is therefore called for.

It is plausible that regression models with skew $t$ error terms
behave quite similarly to analogous cases which employ a regular
$t$ distribution, as for the phenomenon discusses by Fernández \&
Steel (1999). In the numerical work of the next subsection, we
have been driven by considerations described above, and decided to
ignore poles of the log-likelihood very near $\nu=0$. We have
however searched for them, but the only case where we have
successfully located one was with the stackloss data, near
$\nu=0.06$, while the maximum above the threshold $\nu_0=8/13$ was
at $\hat\nu=1.14$.


\subsection{Numerical examples}

\paragraph{AIS data}

It is instructive to examine the outcome of a data fitting process
based on the skew $t$ distribution in a few practical cases. 
Data on several biomedical variables from 202 athletes have been
collected  at the Australian Institute of Sport; see Cook \&
Weisberg (1994) for their description.

\begin{figure}
\psfrag{ais4}[][]{} \psfrag{ais4}[][]{}
\centerline{\includegraphics[width=0.9\hsize,height=9cm]{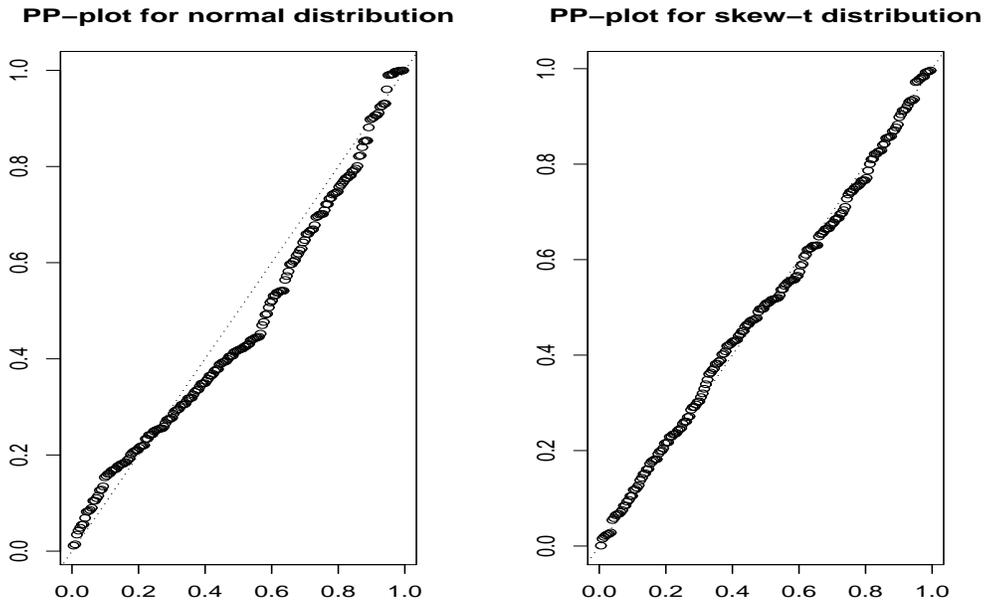}}
\caption{\small\sl AIS data: Healy's plot when either a normal
distribution
    (left-hand side panel)  or a skew t distribution (right-hand side panel
)
    is fitted to the data}
\label{fig:ais-pp}
\end{figure}

We consider here four variables, $\mathit{(BMI, Bfat, ssf, LBM)}$,
which represent represent the body mass index, the percentage of 
body fat, the  sum of skin folds and the lean body mass, respectively.
A $\St_4$ distribution has been fitted to the 202 points, and
Figure~\ref{fig:ais-pp} shows the associated Healy's plot, using
the multivariate normal and the skew $t$ distribution, as
described at the end of Section~\ref{s:LQ}. The plots indicate a
satisfactory fit to the data provided by the skew $t$, markedly
superior to the normal one.

This figure matches with Figure~6 of Azzalini \& Capitanio (1999),
who fit a SN distribution to the same data. While the SN fit was
definitely superior to the normal one, still there was some
discrepancy from the identity line which has now vanished almost
perfectly.

The full list of estimated parameters is not of particular
interest, but it is noteworthy that $\hat\nu=13.7$, which confirms
the presence of somewhat longer tails than the normal
distribution.

We do not present the analogue of Figure~5 of Azzalini \&
Capitanio (1999) because its graphical appearance in our case is
not so markedly different from their Figure~5. These differences
exist, but they become graphically evident only in a summary plot
like the one reported.

\paragraph{Strength of fiber-glass}

Smith \& Naylor (1987) have reported values concerning the
breaking strengths of 1.5\,cm long glass fibers.  These data have
also been considered by Jones \& Faddy (2002) in association with
another form of skew $t$ distribution, and comparison with their
results is the reason for including this example here.

\begin{figure}
\psfrag{glass}[][]{glass-fiber data}
\psfrag{glass}[][]{glass-fiber strength}
\centerline{\includegraphics[width=0.9\hsize,height=9cm]{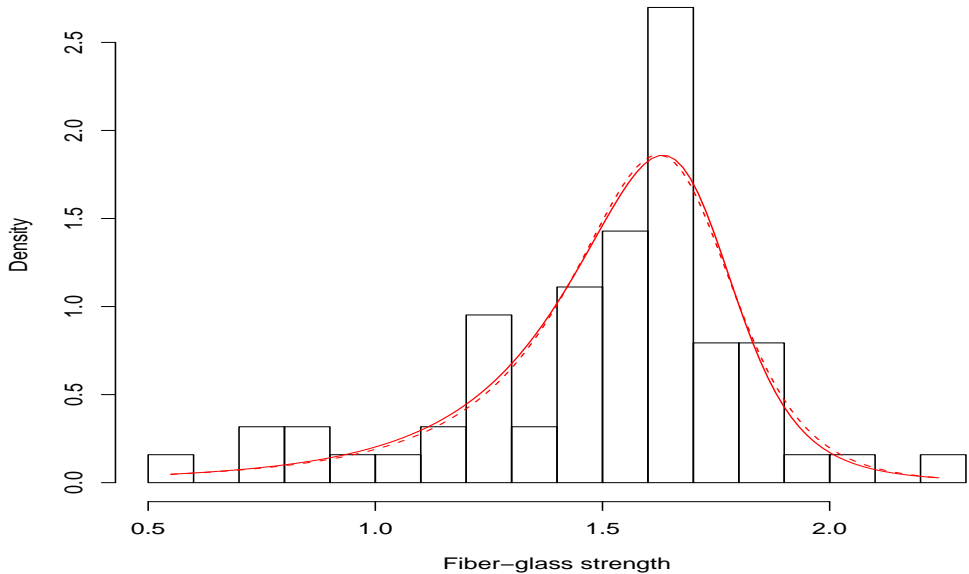}}
\caption{\small\sl Fiber--glass data: histogram and fitted skew
$t$ densities; the continuous curve refers to the density studied
in this paper, the dashed curve refers to Jones' model  }
\label{fig:glass-fit}
\end{figure}

\begin{figure}
\psfrag{glass}[][]{} \psfrag{glass}[][]{}
\centerline{\includegraphics[width=0.9\hsize,height=9cm]{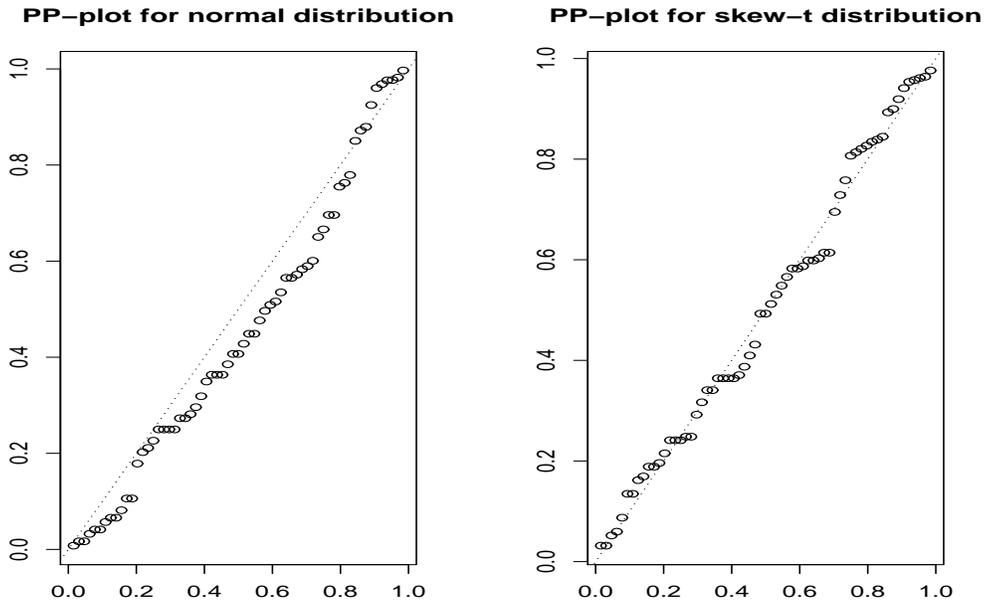}}
\caption{\small\sl Fiber--glass data:
     Healy's plot when either a normal distribution (left panel)
     or a skew t distribution (right panel) is fitted to the glass data}
\label{fig:glass-pp}
\end{figure}

Figure~\ref{fig:glass-fit} shows a histogram of the data and skew
$t$ densities fitted using (\ref{f:pdf-t}) and the Jones'
distribution.  The two parametric densities are graphically very
close, and choice between the two distributions has to be based on
other aspects, rather than empirical adequacy. The Healy plot
associated to (\ref{f:pdf-t}), in Figure~\ref{fig:glass-pp},
confirms  a satisfactory fit of the parametric distribution to the
data.

\begin{figure}
\psfrag{dataset: glass }[][]{} \psfrag{Profile deviance}[][]{}
\centerline{
 \psfrag{alpha}[][]{$\alpha$}
 (a)\includegraphics[width=0.50\hsize,height=9cm]{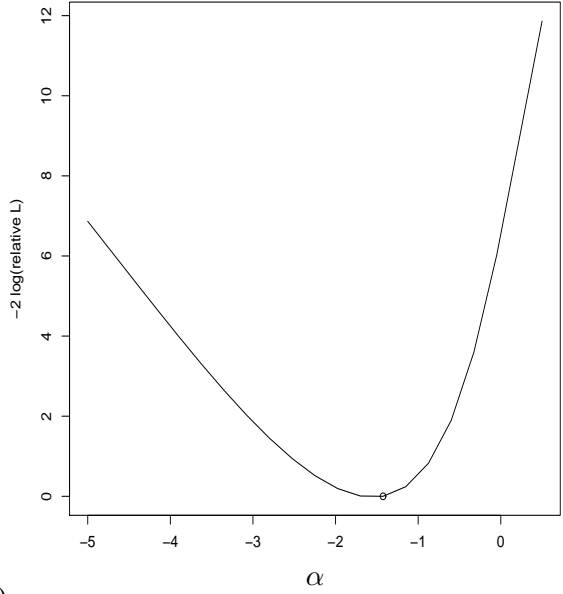}
 ~(b)
 \psfrag{log(df)}[][]{$\log(\nu)$}
 \includegraphics[width=0.50\hsize,height=9cm]{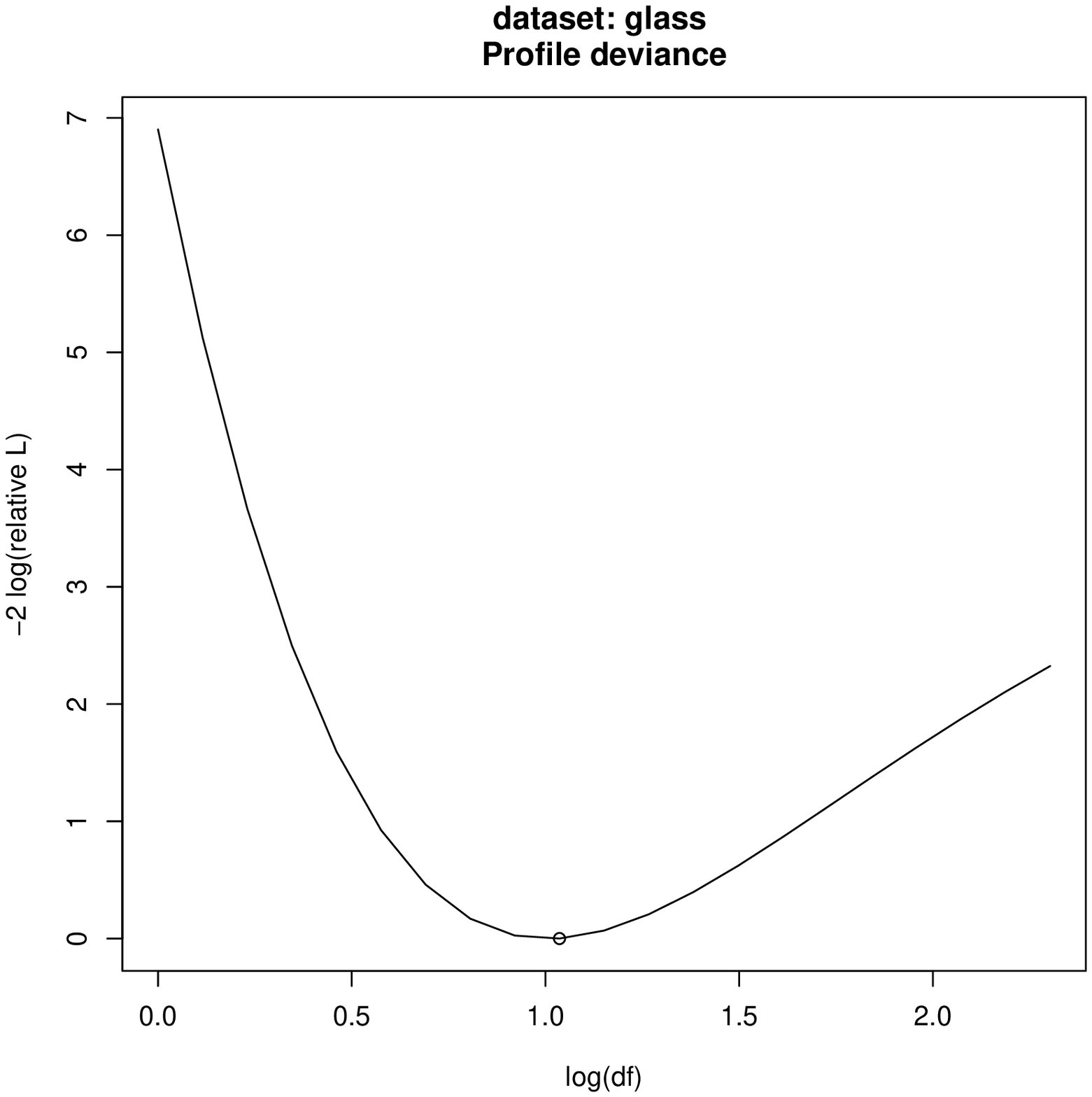}
 }

\centerline{ \psfrag{dataset: glass }[][]{} (c)
 \psfrag{alpha}[][]{$\alpha$}
 \psfrag{log(omega)}[][]{$\log(\omega)$}
 \includegraphics[width=0.50\hsize,height=9cm]{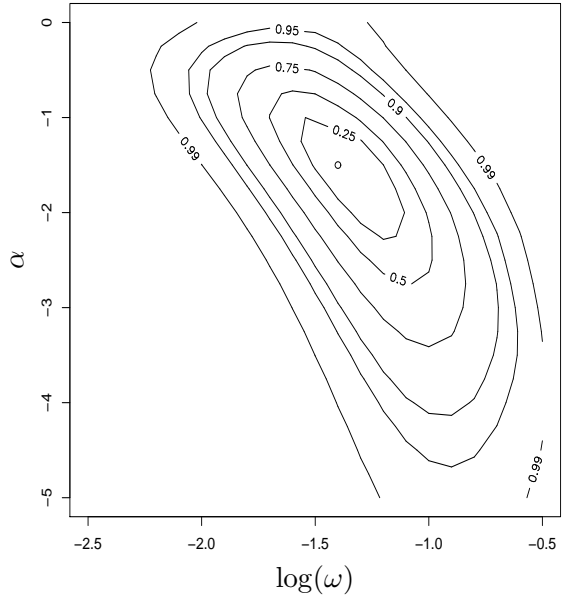}
 ~ (d)
 \psfrag{alpha}[][]{$\alpha$}
 \psfrag{log(df)}[][]{$\log(\nu)$}
 \includegraphics[width=0.50\hsize,height=9cm]{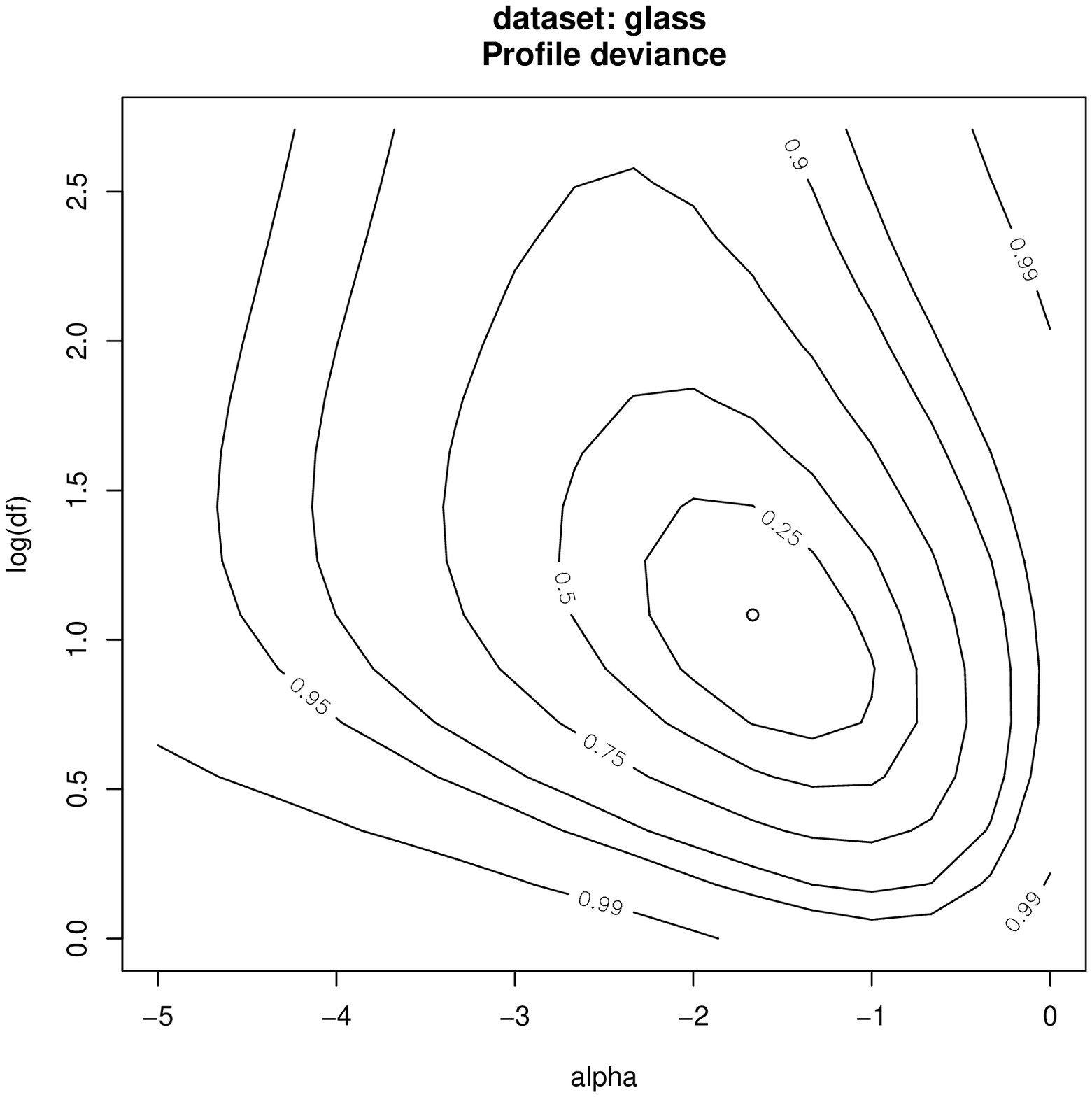}
 } 
\caption{\small\sl Fiber--glass data: twice profile negative
relative log-likelihood for parameters $\alpha$, $\log\nu$,
$(\log\omega,\alpha)$ and $(\alpha,\log\nu)$ are given in panel
(a) to (d), respectively
 respectively}
\label{fig:glass-logL}
\end{figure}

Other interesting features are indicated by twice the profile
log-likelihood functions for the parameters $\alpha$, $\log\nu$,
$(\log\omega,\alpha)$ and $(\alpha,\log\nu)$ reported in panel (a)
to (d) of Figure~\ref{fig:glass-logL}, respectively. The contour
lines for the two parameter cases are chosen to correspond to
differences from the maximum equal to  the quantiles of level
0.50, 0.75, 0.90, 0.95, 0.99 of the $\chi^2_2$ distribution; hence
each contoured region can be interpreted as a confidence region
for the pair of parameters, at the quoted confidence level. As
anticipated earlier, these plots have a quite regular behaviour,
not very far from quadratic functions.

This figure also indicates quite clearly a significant negative
skewness of the  distribution, since the confidence regions up to
level 95\% are entirely on the left of $\alpha=0$. This conclusion
is confirmed by the value of $\hat\alpha$ divided by its standard
error, which is  $-1.55/0.574 \approx -2.70$, with corresponding
$p$-value about 0.7\%. There is also an indication of a long tail
of the distribution, since $\hat\nu=2.73$, but rather higher
values of $\nu$ are not ruled out. These conclusions are broadly
similar to those of Jones \& Faddy (2001); from our analysis there
appears to be a slightly stronger indication of significant
skewness.

\paragraph{Martin Marietta data}

Our next example considers data
taken from Table~1 of Butler, McDonald, Nelson and White (1990).
Based on the arguments presented in that paper,  a linear
regression is introduced
\[  y = \beta_0 + \beta_1 \mathrm{CRSP} + \eps \]
where $y$ is  the excess rate of the Martin Marietta company, CRSP
is an index of  the excess rate of return for the New York market
as a whole and $\eps$ is an error term which in our case is taken
to be distributed as $\St(0,\omega^2,\alpha)$. Data over a period
of $n=60$  consecutive months are available.

\begin{figure}
\centerline{\includegraphics[width=0.8\hsize,height=9.5cm]{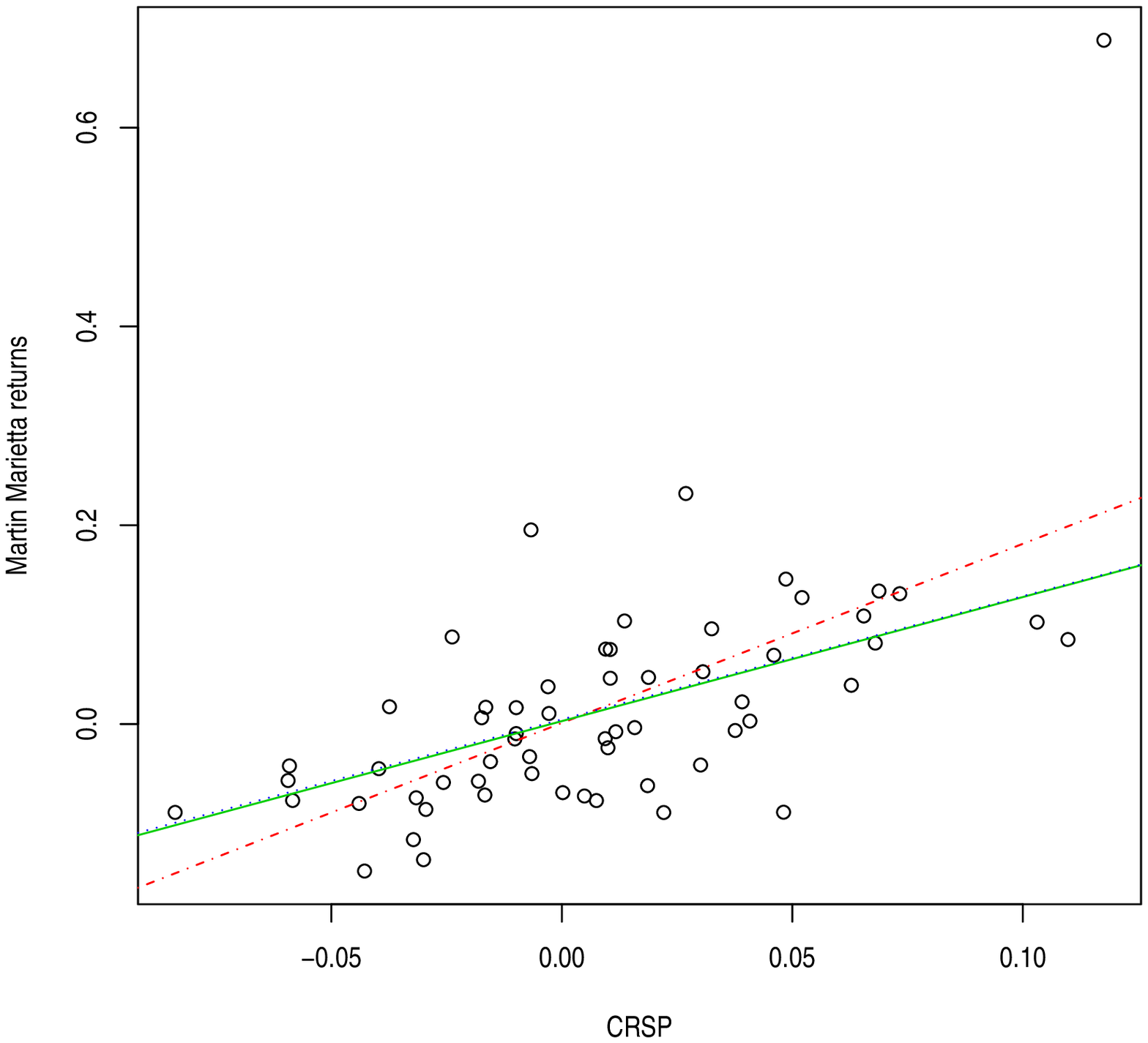}}
\caption{\small\sl Martin Marietta data: scatterplot and fitted
regression lines;
    the dot-dashed line is the least squares fit, the continuous line
    is the one using a skew $t$ error term }
\label{fig:returns-fit}
\end{figure}

The resulting fitted line is shown in Figure~\ref{fig:returns-fit}, 
which displays the scatter-plot of the data with superimposed 
the least squares lines and the line obtained from the above model 
after adjusting for \E{\eps}, whose intercept and slope are
\[  \hat\beta_0+\hat\mathbb{E}\{\eps\}= 0.0029, \qquad 
    \hat\beta_1= 1.248 
\] respectively.
These values are very close to those obtained using the skew $t$ 
distribution of Jones (2001), and the addition of that line to
Figure~\ref{fig:returns-fit} would be barely visible, being
essentially coincident with our line. The estimated skewness
parameter is $\hat\alpha\approx 1.246$ with standardised value
$1.246/0.653\approx 1.908$ and observed significance 5.6\%. The
estimated degrees of freedom are $\hat\nu=3.32 \;(\mathrm{s.e.}
1.43)$.

\begin{figure}
\psfrag{m.marietta}[][]{}
\psfrag{Histogram of res}[][]{}
\centerline{\includegraphics[width=0.9\hsize,height=9cm]{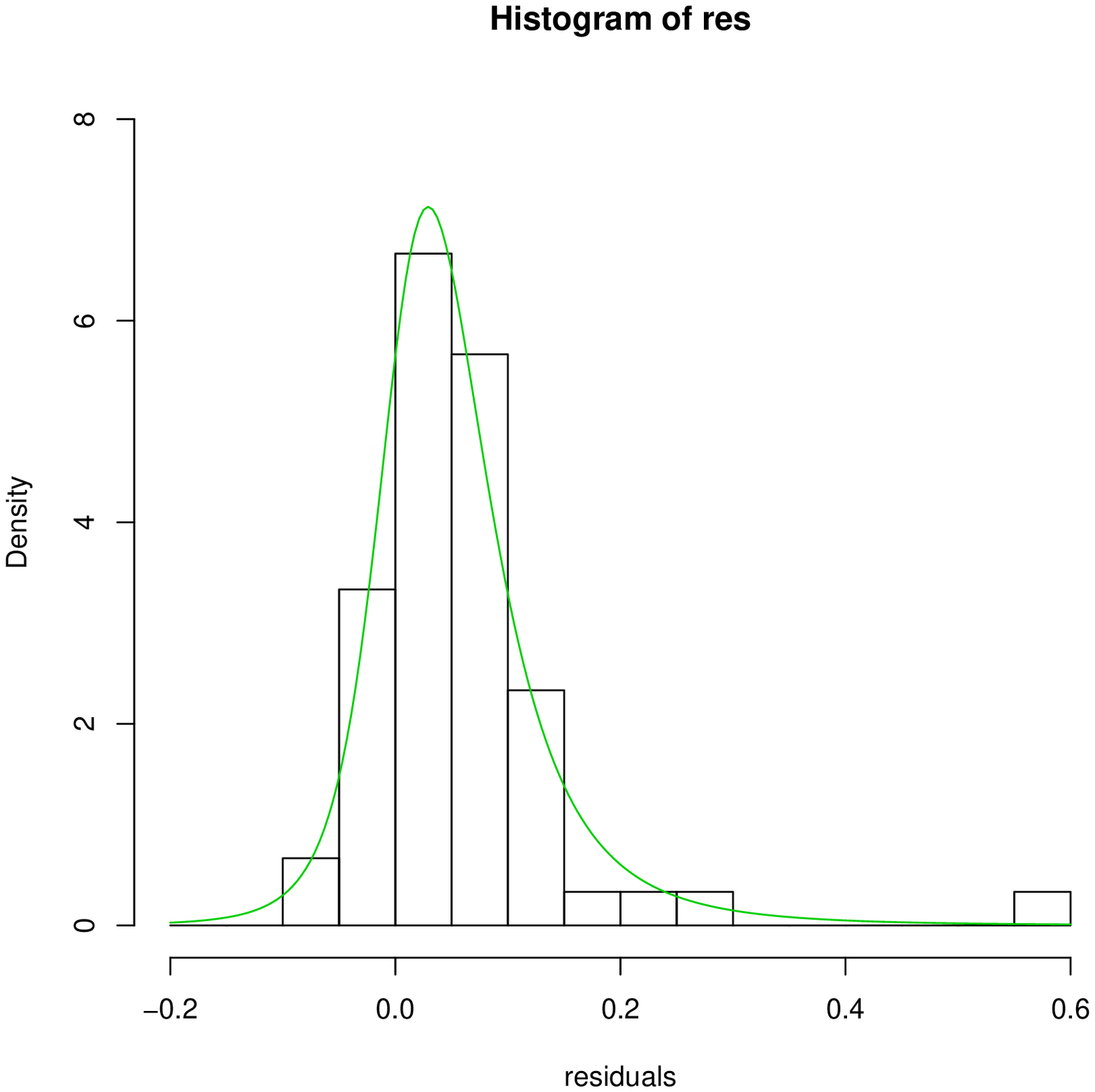}}
\caption{\small\sl Martin Marietta data:
    histogram of the residuals of linear regression and fitted skew $t$
    distribution}
\label{fig:returns-hist}
\end{figure}

\begin{figure}
\psfrag{dataset: m.marietta }[][]{} \psfrag{Profile deviance}[][]{}
\centerline{
 \psfrag{alpha}[][]{$\alpha$}
 \includegraphics[width=0.50\hsize,height=9cm]{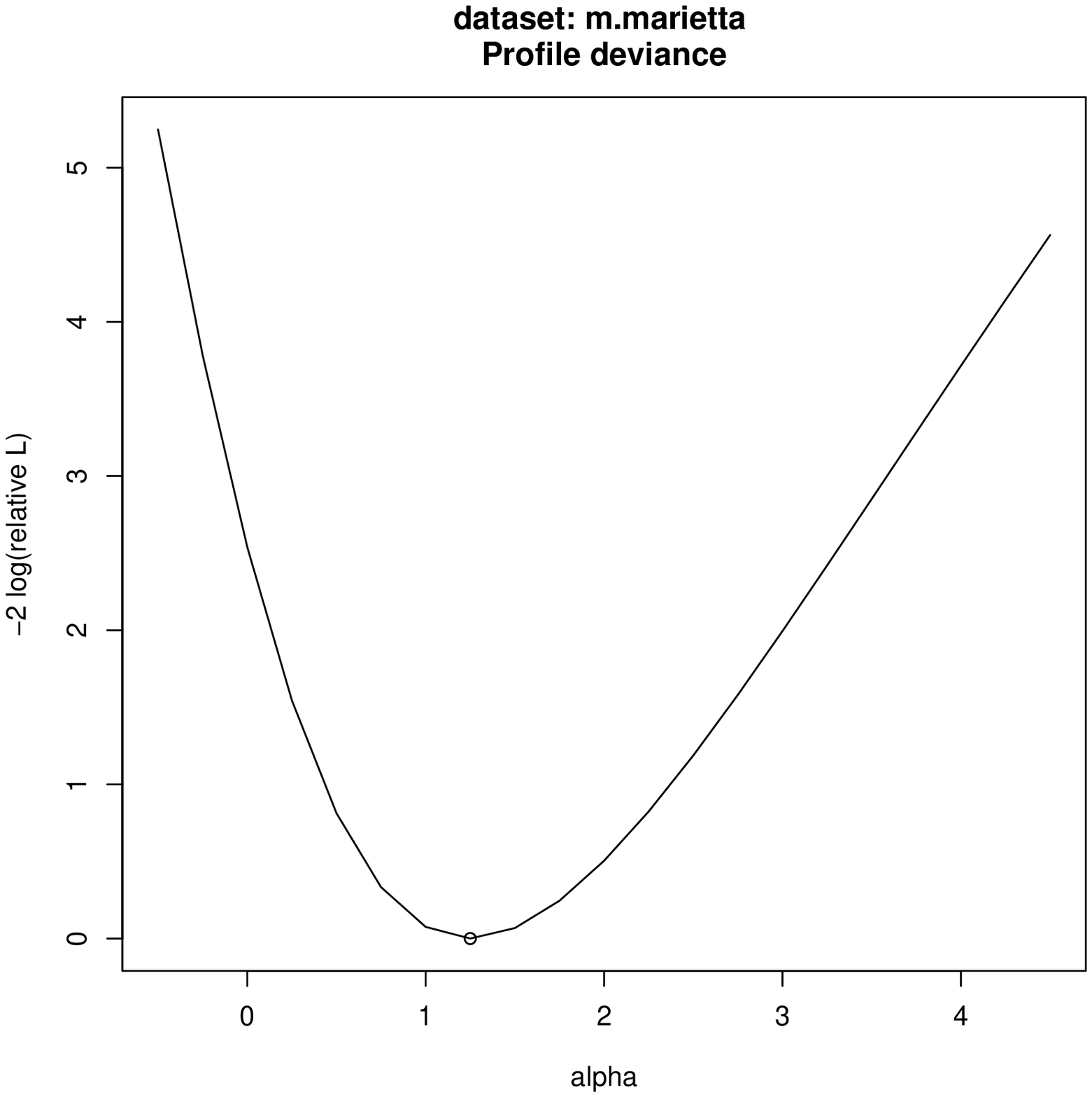}
 ~
 \psfrag{log(df)}[][]{$\log(\nu)$}
 \includegraphics[width=0.50\hsize,height=9cm]{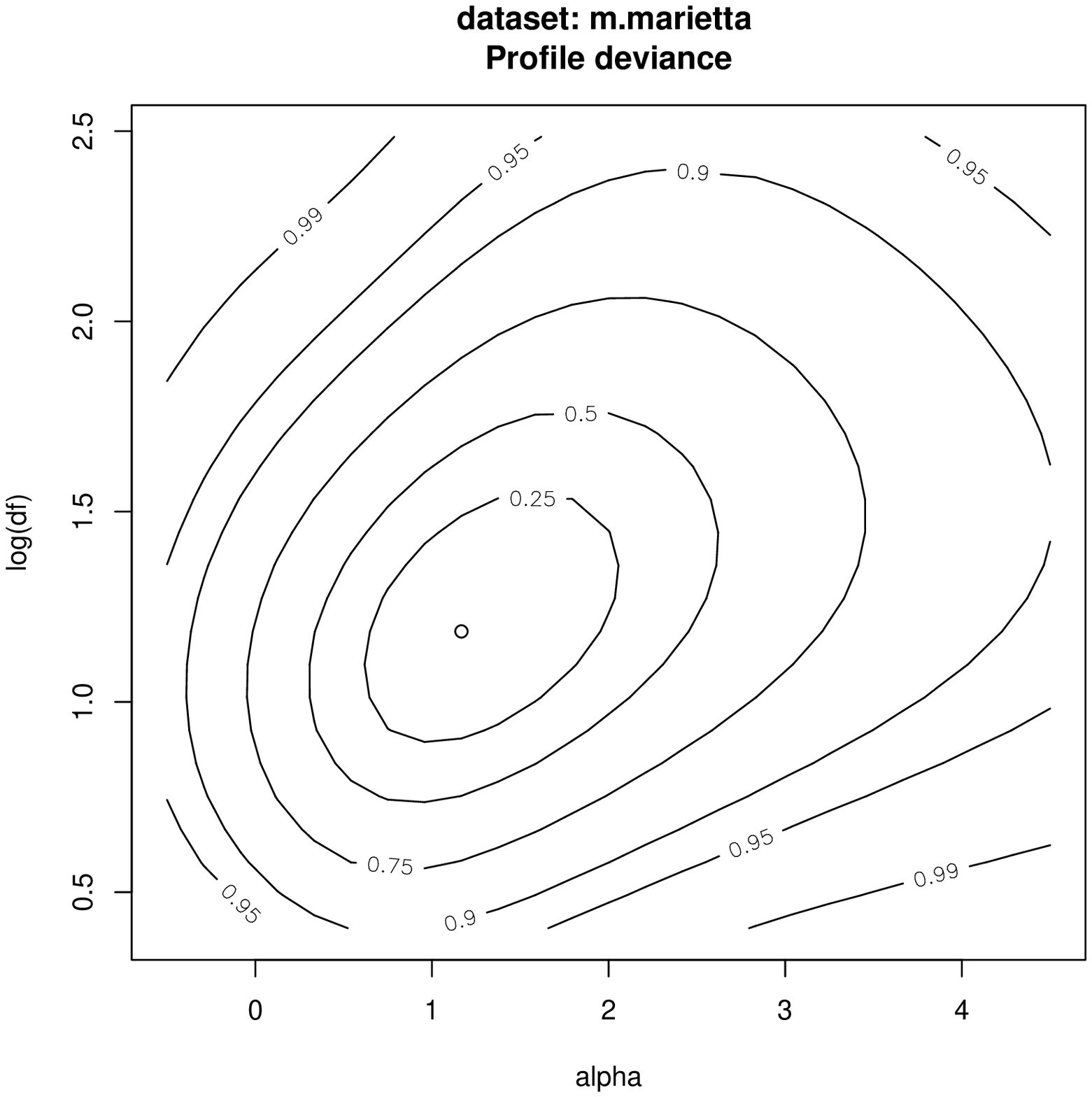}
 }
\caption{\small\sl Martin Marietta data: twice profile negative
  relative log-likelihood for parameters $\alpha$ (left panel) 
  $(\alpha,\log\nu)$ (right panel)}
\label{fig:returns-logL}
\end{figure}

\begin{figure}
\psfrag{m.marietta}[][]{} \psfrag{m.marietta}[][]{}
\centerline{\includegraphics[width=0.9\hsize,height=9cm]{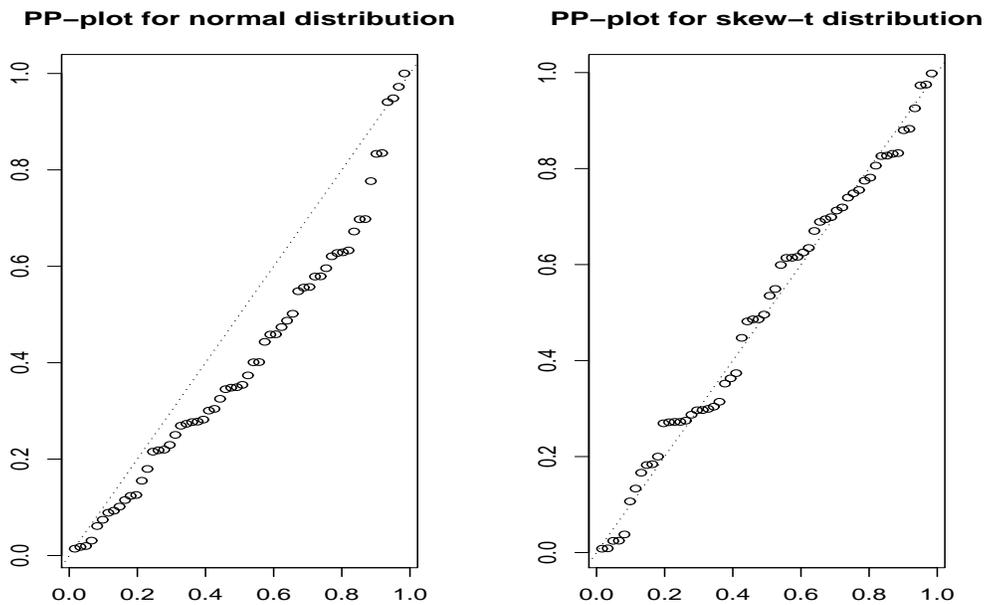}
} \caption{\small\sl Martin Marietta data:
     Healy's plot when either a normal distribution (left panel)
     or a skew t distribution (right panel)}
\label{fig:returns-pp}
\end{figure}

As further indication of the agreement between observed data and
fitted distributions, Figure~\ref{fig:returns-hist} shows the
histogram of the residuals after removing the  line
$\hat\beta_0+\hat\beta_1 \mathrm{CRSP}$, and the fitted skew $t$
density; there appears to be a satisfactory agreement between the
two.
Similarly to Figure~\ref{fig:glass-logL}, the shape of the 
log-likelihood function displayed a nice regular behaviour, as 
indicated by Figure~\ref{fig:returns-logL}.
 Finally, Figure~\ref{fig:returns-pp} compares the Healy's
plots for the normal and a skew $t$ fitted models. Expectedly the
normal model shows obvious inadequacy, while the skew $t$ model
behaves satisfactorily.


\section{Discussion}

A number of broadly related proposals and results have appeared in the
recent literature under the connecting concept of the multivariate
skew normal distribution.  The present paper has examined the
relationships among many of the above proposals, especially of those
dealing with  various formulations of skew elliptical family, by
examining their connections  and providing a more general approach to
obtain several specific results.

Among the broad class of skew elliptical family, the multivariate skew
$t$ distribution offers ample flexibility for adapting itself to a
very wide range of practical situations, and still it maintains
mathematical tractability and a set of appealing formal properties.
Some numerical evidence and the availability of developed software for
inference provide additional support for using the distribution in
practical cases.  
Other interesting distributions have been presented in the literature,
most of which fall under the general umbrella of density (\ref{f:pdf})
and its extensions discussed at the end of Section~\ref{s:c-symmetry}.

A wide and closely interconnected set of specific results is evolving
towards a quite general framework.  Open problems still exists, both
on the probabilistic and on the inferential side of this area of work,
as we have mentioned at various points in the paper, and additional,
yet unexpected results will be discovered.  However, what seems to us
the more important direction of work, at this stage, is to make use
of the available results in tackling real problems. This is the
ultimate test to decide of the actual usefulness of all this work.


\subsection*{Acknowledgments}

We are grateful to Chris Jones for kindly sending us preliminary
versions of his papers, to José Almer Sanqui for drawing our
attention to Roberts paper,  and to an anonymous referee for a
number of helpful comments on an earlier versions of the paper.
This research has been supported by MIUR, Italy, under grant
scheme PRIN 2000.

\clearpage
\appendix
\section*{Appendix}


\section{Proof of Proposition \ref{p:stoch-rep}}

Consider $Y={L\inv}\T (Z-\xi)$, where the $d \times d$ matrix $L$
is such that $\bar\Omega = L\T L$. Then the density of $Y$ is
\[
 2\, f(y;I)G\{w_L(y)\},
\]
where $w_L(-y)=w(-L\T y)=-w_L(y)$. Using the transformation to
spherical coordinates
\[
   Y_j = R\left(\prod_{k=1}^{j-1}\sin\theta_k\right)\cos\theta_j, \quad
   1 \leq j \leq d-1, \qquad
   Y_d = R\left(\prod_{k=1}^{d-2}\sin\theta_k\right)\sin\theta_{d-1},
\]
where $R>0$, $\theta_k \in [0,\pi)$, for $k=1,\ldots,d-2$ and
$\theta_{d-1}\in [0,2\pi)$, and taking into account that the
Jacobian is $r^{d-1}\prod_{k=1}^{d-2}(\sin\theta_k)^{d-k-1}$, we
have
\begin{eqnarray*}
 && f_{\theta,R}(\mathbf{\theta},r) \:= \\
 &&\quad =   2\,c_d \gen (r^2)r^{d-1}\prod_{k=1}^{d-2}(\sin\theta_k)^{d-k-1}
       G\{w_L(r\cos\theta_1,r\sin\theta_1\cos\theta_2,\ldots,
       r\sin\theta_1\ldots \sin\theta_{d-1})\}\\
 &&\quad = 2\,c_d \gen (r^2)r^{d-1}\prod_{k=1}^{d-2}(\sin\theta_k)^{d-k-1}
     G\{w_L^*(\theta,r)\}
\end{eqnarray*}
where $\mathbf{\theta} = (\theta_1,\ldots,\theta_{d-1})\T$, and
$w^*_L(\theta,r)=
w_L(r\cos\theta_1,r\sin\theta_1\cos\theta_2,\ldots,
   r\sin\theta_1\ldots \sin\theta_{d-1})$. Notice that
$\dfrac{2\pi^{d/2}}{\Gamma(d/2)}c_d \gen (r^2)r^{d-1}$ is the
density of the radius in the stochastic representation
(\ref{f:ru}) of the elliptical random vector $U$, say, having
density $f$, and
$\dfrac{\Gamma(d/2)}{2\pi^{d/2}}\prod_{k=1}^{d-2}(\sin\theta_k)^{d-k-1}$
is the spherical coordinates representation of the uniform
distribution on the unit sphere of $\Real^d$; see Fang \emph{et
al.} (1990, Section 2.2.3). From Proposition \ref{th:transf} it
follows that $R^2 \equald Y\T Y \equald U\T U$, so that the
marginal density of $R$ is given by
\[
f_R (r)= \dfrac{2\pi^{d/2}}{\Gamma(d/2)}c_d \gen (r^2)r^{d-1}.
\]
By integrating the joint density $f_{\theta,R}$ with respect to
$r$, the marginal density of $\theta$ turns out to be
\begin{eqnarray*}
   f_{\theta}(\mathbf{\theta})
               &=& \frac{\Gamma(d/2)}{2\pi^{d/2}}
                    \prod_{k=1}^{d-2}(\sin\theta_k)^{d-k-1} \,
                    2\int_0^\infty  f_r(r)G\{w_L^*(\theta,r)\}\d r \\
                &=&  \frac{\Gamma(d/2)}{\pi^{d/2}}
                    \prod_{k=1}^{d-2}(\sin\theta_k)^{d-k-1}
                    \pr{X\leq   w_L^*(\theta,R)},
\end{eqnarray*}
where $X$ is a random variable with cumulative distribution
function $G$.

The conditional density of $\theta$ given $R=r$ is equal to
\[
f_{\theta|R=r}(\theta) =
      \dfrac{\Gamma(d/2)}{\pi^{d/2}}\prod_{k=1}^{d-2}(\sin\theta_k)^{d-k-1}
      \:G\{w_L^*(\theta,r)\},
\]
which is a density of type (\ref{f:pdf}) with location parameter
$(\pi/2,\ldots,\pi/2,\pi)$. In fact for any $r>0$ and any matrix
$L$ the equality
\[
   w_L^*(\pi-\theta_1,\pi-\theta_2,\ldots,\pi+\theta_{d-1},r)
     = -w_L^*(\theta_1,\theta_2,\ldots,\theta_{d-1},r)
\]
holds true, and consequently the random variable $W_L^*=
w_L^*(\theta,r)$ is symmetrically distributed around $\pi$. Then,
using Lemma 1 in Azzalini \& Capitanio (1999, p.\,599), the result
follows.~QED


\section{Derivatives of the skew $t$ log-likelihood}

Write $U=(u_1,\ldots,u_n)\T$. Then the derivatives of
(\ref{f:logL-t}) are obtained from
\begin{eqnarray*} 
\frac{\partial \ell}{\partial \beta }  &=&
    -2\, X\T \diag(\tilde{g}_Q+\tilde{T}_1\odot \dot{t}_Q) U  \Omega\inv
    - X\T \diag(\tilde{T}_1 \odot \dot{t}_L) 1_n \eta\T
\\
\frac{\partial \ell}{\partial A }  &=&
    2 \:\mbox{upper triangle of}\left(
   \:D\: A\: U\T\: \diag(\tilde{g}_Q + \tilde{T}_1 \odot \dot{t}_Q)\: U\right)
\\
\frac{\partial \ell}{\partial D }  &=&
    I_d \odot \left(
       A\: U\T\diag(\tilde{g}_Q + \tilde{T}_1 \odot \dot{t}_Q) U\:A\T\right)
     + \half n D\inv
\\
\frac{\partial \ell}{\partial \eta}  &=&
    U\T \diag(\tilde{T}_1 \odot \dot{t}_L) 1_n
\\
\frac{\partial \ell}{\partial \nu }  &=&
    \sum \left(\frac{\partial \log g_d}{\partial \nu}
         + \frac{\partial \log T_1(t;\nu+d)}{\partial\nu}   \right)
\\
%
\end{eqnarray*}
where the components of the vectors are  obtained by evaluation of
the quoted expressions at each of the $n$ observations, $\odot$
denotes the Hadamard (or element-wise) product and
\begin{eqnarray*}
\tilde{g}_Q &=& \partial{\,\log g_d(Q;\nu)}/\partial{Q}
           = -\frac{\nu+d}{2\nu}\left(1+Q/\nu\right)\inv\\
\tilde{T}_1 &=& \partial{\,\log T_1(t;\nu+d)}/\partial{t}
           = T_1(t;\nu+d)\inv t_1(t;\nu+d)\\
\dot{t}_L &=& \partial{t(L,Q,\nu)}/\partial{L}
           = \radice{\frac{\nu+d}{Q+\nu}}\\
\dot{t}_Q &=& \partial{t(L,Q,\nu)}/\partial{Q}
           = -\frac{L \radice{\nu+d}}{2(Q+\nu)^{3/2}} \\
\frac{\partial \log g_d}{\partial \nu}
          &=& \half\left(\psi(\half(\nu+d)) - \psi(\half\nu) - d/\nu
             +\frac{(\nu+d)Q}{\nu^2(1+Q/\nu)}  - \log(1+Q/\nu) \right) \\
\end{eqnarray*}
denoting by $\psi$ the digamma function. What is not given above
is an expression for
\[
   \frac{\partial \log T_1(t(L,Q,\nu); \nu+d)}{\partial\nu}
\]
which appears intractable and  must be evaluated numerically.

For transforming  the above derivatives of $D$ and $\nu$ into
those of their logarithmic transform, we just use the chain rule
\[
  \frac{\partial \ell}{\partial \rho}
          =  \frac{\partial \ell}{\partial D^*}\:(-2 D^*)\,,
\qquad
  \frac{\partial \ell}{\partial \log \nu}
          =  \frac{\partial \ell}{\partial \nu}\:\nu
\]
where $D^*$ denotes the diagonal of $D$.

The above expressions do not lend themselves to further
differentiation. Therefore, in the numerical work described in
Section~\ref{s:t-stat}, the observed information matrix has been
obtained via numerical differentiation of the first derivatives.


\subsection*{References}

\biblioitem
  Arnold, B.C.\ and Beaver, R.J. (2000a). Hidden truncation models.
  \emph{Sankhy\={a}} \textbf{62}, 22--35.

\biblioitem
  Arnold, B.C.\ and Beaver, R.J. (2000b). The skew Cauchy distribution.
  \emph{Statist.\,Prob.\,Lett.} \textbf{49}, 285--290.

\biblioitem
  Arnold, B.C.\ and Beaver, R.J. (2002). Skewed multivariate models related
  to hidden truncation and/or selective reporting.
  \emph{Test} \textbf{11}, 7--54.

\biblioitem
  Azzalini, A. (1985). A class of distribution which includes the normal
  ones.   \emph{Scand. J. Statist.} {\bf 12}, 171--8.

\biblioitem
  Azzalini, A. and Capitanio, A. (1999).  Statistical
  applications of the multivariate skew normal distribution.  \emph{J.
  Roy. Statist. Soc., B} \textbf{61} 579--602.

\biblioitem
  Azzalini, A. and Dalla Valle, A. (1996).  The multivariate
  skew normal distribution.  \emph{Biometrika} \textbf{83}, 715--26.

\biblioitem
  Branco, M.~D.\ and Dey, D.~K.\ (2001).
  A general class of multivariate skew elliptical distributions.
  \emph{Journal of Multivariate Analysis} \textbf{79}, 99--113.

\biblioitem
  Brownlee, K. A. (1960, 2nd ed.\ 1965).
  \emph{Statistical Theory and Methodology in Science and Engineering}.
  New York: Wiley. 

\biblioitem
  Butler, R. L., McDonald,J. B., Nelson, R. D. and White, S. B. (1990).
  Robust and partly adaptive estimation of regression models.
  \emph{Rev. Econ. Statist.} \textbf{72}, 321--327.

\biblioitem
  Capitanio, A., Azzalini, A. and Stanghellini, E. (2003).
  Graphical models for skew normal variates.
 \emph{Scand.\,J.\ Statist.} \textbf{30}, 129--144.

\biblioitem 
  Cook, R.\,D. and Weisberg, S. (1994).
  {\em An Introduction to Regression Graphics}. Wiley,  New York.

\biblioitem 
  David, H.\,A. (1981). \emph{Order statistics}, 2nd edition.
  Wiley, New York.

\biblioitem 
  Fang, K.-T., Kotz, S. and  Ng, K. (1990).
  \emph{Symmetric multivariate and related distributions}.
  Chapman \& Hall, London.

\biblioitem
  Fernández, C. and Steel, M.\ F.\ J.\  (1998).
  On Bayesian modelling of fat tails and skewness.
  \emph{J.~Am.\ Statist.\ Assoc.} \textbf{93}, 359--371.

\biblioitem
  Fernández, C. and Steel, M.\ F.\ J.\  (1999).
  Multivariate Student-$t$ regression models: pitfalls and inference.
  \emph{Biometrika} \textbf{86}, 153--168.

\biblioitem
  Genton, M. G.,  He, L. and Liu, X. (2001).
  Moments of skew normal random vectors  and  their quadratic forms.
  \emph{Statist. \& Prob.\ Lett.} \textbf{51}, 319--325.

\biblioitem
Genton, M. G., \& Loperfido, N. (2002).
  Generalized skew-elliptical distributions and their quadratic forms.
  Institute of Statistics Mimeo Series No.\,2539,
  North Carolina State University. \\
  \texttt{http://www.stat.ncsu.edu/library/mimeo.html}

\biblioitem
  Genz, A. and Bretz, F. (1999).
   Numerical Computation of Multivariate $t$-Probabilities with Application
    to Power Calculation of Multiple Contrasts.
   \emph{J. Stat. Comp. Simul.} \textbf{63}, 361--378.

\biblioitem
  Gupta, A.K., Gonzáles-Farías, G. and Domínguez-Molina, J.\,A. (2001).
  A Multivariate Skew Normal Distribution. Report I-01-19,
  \texttt{http://www.cimat.mx/reportes}

\biblioitem
  Healy, M. J. R. (1968). Multivariate normal plotting.
  \emph{Appl. Statist.} {\bf17}, 157--161.


\biblioitem
  Jones, M.\,C. (2002). Multivariate $t$ and Beta distributions
  associated with the multivariate $F$ distributions.
  \emph{Metrika}, \textbf{54}, {215--231}.

\biblioitem
  Jones, M.C. (2001).  A skew $t$ distribution.
  In \emph{Probability and Statistical Models with Applications:
  a Volume in Honor of Theophilos Cacoullos},
  eds: C.\,A. Charalambides, M.\,V. Koutras and N.\ Balakrishnan.
  Chapman and Hall, London, 269--278.

\biblioitem
  Jones, M.C. and Faddy, M.J.(2001).
  A skew extension of the $t$ distribution, with applications.
  To appear.

\biblioitem
  Kano, Y. (1994).
  Consistency property of the elliptic probability density functions.
  \emph{J.\ Multiv.\ An.\ } \textbf{51}, 139--147.


\biblioitem
   Loperfido, N. (2001).
   Quadratic forms of skew normal random vectors.
   \emph{Statistics \& Probability Letters} \textbf{54}, 381--387.

\biblioitem
   Loperfido, N. (2002).
   Statistical implications of selectively reported inferential results.
   \emph{Statistics \& Probability Letters} \textbf{56}, 13--22.


\biblioitem
   Roberts, C. (1966).
   A correlation model useful in the study of twins.
   \emph{J. Am. Statist. Assoc.} \textbf{61},  1184--1190.

\biblioitem
   Sahu, S.\,K., Dey, D.\,K.\ and Branco, M. (2001).
   A New Class of Multivariate Skew Distributions with Applications
   to Bayesian Regression Models.
   Tech. report\newline \texttt{
   http://www.maths.soton.ac.uk/staff/Sahu/research/papers/skew.html}

\biblioitem
  Smith, R. L. and Naylor, J. C. (1987).
  A comparison of maximum likelihood and Bayesian estimators for the
  three-parameter Weibull distribution.
  \emph{Appl.\ Statist.}, \textbf{36}, 358--369.

\biblioitem
  Szab{\l}owski, P.\ J.\ (1998).
  Uniform distributions on spheres in finite-dimensional $L_\alpha$
  and their generalization.
   \emph{Journal of Multivariate Analysis} \textbf{64}, 103-117.

\biblioitem
   Zuo, Y., and Serfling, R. (2000).
   On the performance of some robust nonparametric location measures
   relative to a general notion of multivariate symmetry.
   \emph{J.\ Statistical Planning and Inference} \textbf{84}, 55--79.

\end{document}

%% file: aa-macros.tex
\usepackage{amsfonts}
\newcommand{\anymode}[1]{\ifmmode{#1}\else\mbox{$#1$}\fi}
\newcommand{\biblioitem}[1]{\par \frenchspacing
           \vbox{\noindent \hangindent=2em \hangafter=1{#1}}
           \vskip 2ex}
\def\biblioitem{\par \vskip 1ex \noindent \hangindent=2em \hangafter=1}

\newcommand{\NB}[1]{\texttt{/\textsuperscript{NB}}%
          \marginpar{\raggedright\sl\small#1\hfill}}	  

\long\def\Nota#1{\footnote{#1}\kern-0.2em\NB{cfr Nota$^{(\thefootnote)}$}}
\def\linea#1{\ifhmode\hfill\break\fi\hbox to \hsize{#1}}

\newcommand{\inv}{^{-1}}

\def\vc{v\kern -0.1em .c.\relax}

\newcommand{\dfrac}[2]{\displaystyle{\frac{#1}{#2}}}
\newcommand{\diag}{\mbox{\rm diag}}
\newcommand{\equald}{\stackrel{d}{=}}
\newcommand{\E}[2][]{
   \ensuremath{\mathbb{E}_{#1}\!\left\{\displaystyle{#2}\right\}}}
\newcommand{\eps}{\varepsilon}

\newcommand{\half}{\mbox{$\textstyle \frac{1}{2}$}}   
\long\def\ignore#1{}

\newcommand{\indep}{\perp\kern-0.5em\perp}

\newcommand{\pr}[2][]{
   \ensuremath{\mathbb{P}_{#1}\!\left\{\displaystyle{#2}\right\}}}
\newcommand{\radice}[1]{\left(#1\right)^{1/2}}
\newcommand{\recradice}[1]{\left(#1\right)^{-1/2}}

\newcommand{\Real}{\mathbb{R}}

\newcommand{\T}{^{\top}}
\newcommand{\var}[2][]{
   \ensuremath{\textrm{var}_{#1}\!\left\{\displaystyle{#2}\right\}}}

\newtheorem{theorem}{Theorem}
\newtheorem{lemma}[theorem]{Lemma}

\newtheorem{proposition}[theorem]{Proposition}
\newcommand{\SN}{\mathrm{SN}{}}
\newcommand{\N}{\mathrm{N}{}}

\renewcommand{\d}{\mathrm{d}}

